\documentclass[a4paper,fleqn,usenatbib]{mnras}
\usepackage{color}
\usepackage{natbib}
\usepackage{paralist}
\usepackage{graphicx}
\usepackage{epstopdf}
\usepackage{epsfig}
\usepackage{amsmath}
\usepackage{tablefootnote}

\usepackage{graphicx}   
\usepackage{amsmath}    
\usepackage{amssymb}    
\usepackage{multicol}        
\usepackage{bm}         
\usepackage{pdflscape}  

\newcommand\teff{{T_{\rm eff}}}

\newcommand\lta{\mathrel{\hbox{\raise 0.6 ex \hbox{$<$}\kern
                   -1.8 ex\lower .5 ex\hbox{$\sim$}}}}
\newcommand\gta{\mathrel{\hbox{\raise 0.6 ex \hbox{$>$}\kern
                   -1.7 ex\lower .5 ex\hbox{$\sim$}}}}
\newcommand{\bea}{\begin{eqnarray}}
\newcommand{\eea}{\end{eqnarray}}

\newcommand\species[2]{#1 {\sc #2}}
\newcommand\iso[2]{$^{\rm #1}$#2}

\def\ciso{{$^{12}$C/$^{13}$C}}

\def\teff{\mbox{$T_{\rm eff}$}}
\def\logg{\mbox{log~{\it g}}}
\def\vmicro{\mbox{$\xi_{\rm t}$}}
\def\kmsec{\mbox{km~s$^{\rm -1}$}} 
                                          
\def\eg{\mbox{e.g.}}                                          
\def\loggf{\mbox{log~{\it gf}}}

\def\ncap{\mbox{\textit{$n$}-capture}}                        
\def\loggf{\mbox{log~{\it gf}}}                               
                                                              
\usepackage[T1]{fontenc}                                      
\usepackage{ae,aecompl}                                       
                                                              
\usepackage{newtxtext,newtxmath}                              
                                                              
\title[IR abundances of the RGs in the NGC 752]{\textit{Chemical Abundances Of Open
Clusters From High-Resolution Infrared Spectra. II. NGC 752}} 
                                                              
\author[G. B\"{o}cek Topcu  et al.]{                          
        G. B\"{o}cek Topcu$^{1}$\thanks{Contact e-mail:       
           gamzebocek@gmail.com (GBT);                        
        melike.afsar@ege.edu.tr (MA); chris@verdi.as.utexas.edu (CS);        
        cpilacho@indiana.edu (CAP); pavelden@uvic.ca (PAD);   
        vandenbe@uvic.ca (DAV);                               
        gmace@astro.as.utexas.edu (GNM);                      
        hkim@gemini.edu (HK);                                 
        ksokal@utexas.edu (KS);                               
        dtj@astro.as.utexas.edu (DTJ)},                       
        M. Af\c{s}ar$^{1,2}$,                                 
        C. Sneden$^{2}$,                                      
        C. A. Pilachowski$^{3}$,
\newauthor
        P. A. Denissenkov$^{4}$,                              
        D. A. VandenBerg$^{4}$,                               
        D. Wright$^{2}$,                                      
        G. N. Mace$^{2}$,                                     
\newauthor
        D. T. Jaffe$^{2}$,
        E. Strickland$^{2}$,                                  
        H. Kim$^{5}$, and
        K. R. Sokal$^{2}$ \\
$^{1}$Department of Astronomy and Space Sciences,             
                 Ege University, 35100 Bornova, \.{I}zmir, Turkey \\         
$^{2}$Department of Astronomy and McDonald Observatory,       
                 The University of Texas, Austin, TX 78712 \\ 
$^{3}$Indiana University, Department of Astronomy SW319, 727 E 3rd Street,   
                 Bloomington, IN 47405 USA \\                 
$^{4}$Department of Physics and Astronomy, University of Victoria,
                 Victoria, BC, V8W 2Y2, Canada            \\                 
$^{5}$Gemini Observatory, Casilla 603, La Serena, Chile}

\date{Accepted 2019 October 18}                                                       
                                                              
\pubyear{2019}                                                
                                                              
\begin{document}                                              
\label{firstpage}                                             
\pagerange{\pageref{firstpage}--\pageref{lastpage}}           
\maketitle

\begin{abstract}                                              

We present a detailed near-infrared chemical abundance 
analysis of 10 red giant members of the Galactic open cluster NGC 752. 
High-resolution (R$\simeq$45000) near-infrared spectral data were 
gathered with the Immersion Grating Infrared Spectrograph (IGRINS), 
providing simultaneous coverage of the complete $H$ and $K$ bands.
We derived the abundances of H-burning (C, N, O), $\alpha$ (Mg, Si, S, Ca), 
light odd-Z (Na, Al, P, K), Fe-group (Sc, Ti, Cr, Fe, Co, Ni) and 
neutron-capture (Ce, Nd, Yb) elements.  
We report the abundances of S, P, K, Ce, and Yb in NGC~752 for the first time.
Our analysis yields solar metallicity and solar abundance
ratios for almost all of the elements heavier than the CNO group in 
NGC~752.
O and N abundances were measured from a number of OH and CN features 
in the $H$ band, and C abundances were determined mainly from 
CO molecular lines in the $K$ band. 
High excitation \ion{C}{i} lines present in both near-infrared and optical 
spectra were also included in the C abundance determinations.
Carbon isotopic ratios were derived from the R-branch band heads of 
first overtone (2$-$0) and (3$-$1) $^{12}$CO and (2$-$0) $^{13}$CO lines near 
23440 \AA\ and (3$-$1) $^{13}$CO lines at about 23730 \AA. 
The CNO abundances and \ciso\ ratios are all consistent with our giants
having completed ``first dredge-up'' envelope mixing of CN-cyle products.
We independently assessed NGC~752 stellar membership from Gaia astrometry,
leading to a new color-magnitude diagram for this cluster.
Applications of Victoria isochrones and MESA models to these data yield
an updated NGC~752 cluster age (1.52~Gyr)
and evolutionary stage indications for the program stars.
The photometric evidence and spectroscopic light element abundances all
suggest that the most, perhaps all of the program stars are members of the
helium-burning red clump in this cluster.
   
\end{abstract}                                                
                                                              
\begin{keywords}                                              
stars: abundances -- stars:  atmospheres. Galaxy: open clusters and          
associations: individual: NGC 752                             
\end{keywords}


\section{Introduction}\label{intro}

Open star clusters provide important snapshots of the chemistry of the
Galactic disk with time because they can be photometrically tagged with 
ages, which are difficult to assess for individual field stars.
Most open clusters (OCs) are young, $t$~$\leq$~2~Gyr.
Very few of them are truly old, $t$~$>$~7~Gyr; only Be~17 and NGC~6791
appear to have ages approaching 10~Gyr 
(\eg, \citealt{salaris04}, \citealt{brogaard12}).
Fortunately there are many intermediate-age clusters close enough to the
Sun that chemical compositions of their brighter members can be studied 
at high spectroscopic resolution. 
The recent Gaia DR2 catalog now provides opportunities for more accurate 
membership and evolutionary state data for OC red giant members.
Several groups are conducting extensive OC abundance studies 
using echelle spectrographs in the optical spectral region.

Optical spectroscopy of OCs has a fundamental observational limit caused
by Galactic disk dust extinction.
While more than 1000 OCs have been cataloged, most are too obscured to
yield detailed information at optical wavelengths.
This problem is especially acute for clusters at small Galactocentric radii.
High-resolution spectroscopy at more transparent infrared wavelengths
($IR$, $\lambda$~$\geq$~1~$\mu$m) is essential for further progress in 
OC chemical composition studies.

We have begun a program to use $H$ and $K$ band high-resolution 
spectroscopy to determine reliable metallicities and elemental abundance 
ratios for OCs spanning a large range of Galactocentric distances.
Special emphasis is put on determining accurate abundances for
the CNO group and other light elements, which have many transitions
in these $IR$ bands.
As steps toward this goal we first are performing combined optical/infrared
spectroscopic analyses of three relatively nearby and well-studied 
intermediate-age OCs that suffer only small amounts of interstellar dust 
extinction:  NGC~6940, NGC~752, and M67.
Our $IR$ spectra are obtained with the Immersion Grating Infrared Spectrograph 
(IGRINS), which offers complete $H$ and $K$ spectral coverage 
(1.45$-$2.5~$\mu$m), 
with a band gap of only 0.01~$\mu$m lost instrumentally between
the two bands (the gap due to telluric absorption is $\sim$0.2~$\mu$m).
IGRINS delivers high spectral resolution similar to those of most 
optical spectrographs used for abundance analysis.  

For NGC~6940, \cite{bocek19}, hereafter Paper~1, reported analyses of IGRINS 
spectra of 12 red giant members, the same stars with atmospheric parameters 
and detailed abundance sets derived from high-resolution optical spectra 
by \cite{bocek16} (hereafter BT16).
Among their principal results were:  (a) good agreement in all cases of 
optical and $IR$ abundances; (b) determination for the first time 
in NGC~6940 the abundances of S, P, K, Ce, and Yb, and much strengthened
abundances of Mg, Al, Si, and Ca; (c) derivation of much more reliable 
abundances of the CNO group; (d) discovery of one star with evidence of
high-temperature proton fusion products; and (e) improved assessment of
the NGC~6940 color-magnitude diagram, with clear assignment of most of
the program stars to the He-burning red clump.

In this paper we report results of a similar optical/IR study for 
10 red giant (RG) members of NGC~752.
This relatively nearby OC has been the subject of several abundance
studies (\eg, \citealt{pilachowski88,carrera11,reddy12}).
Recently \cite{lum19} have presented an extensive large-sample analysis 
of 6 giant and 23 main-sequence stars.
Our optical spectroscopic investigation was published in \cite{bocek15} 
(hereafter BT15).
In this study we focus especially on the CNO abundances 
and \ciso\ ratios, interpreting the results within a more complete wavelength 
window. 
This larger spectral coverage leads us to a better analysis of the evolutionary
status of the RG members.
In addition to the red giant abundances derived from IGRINS spectra,
we have revisited the questions of the distance and age of NGC~752 with 
\textit{Gaia} DR2 data, and have re-considered the evolutionary states of these stars 
via new stellar isochrone computations.
The Gaia kinematic and photometric data leading to NGC~752 membership,
distance, and age are discussed in \S\ref{gaia}.
In \S\ref{obs} we summarize the IGRINS observations and reductions. 
The methods used to derive the chemical abundances and the 
temperatures of the target stars are described in \S\ref{modopt} and 
\S\ref{ldrcomp}, respectively, while the results of the abundance 
determinations are given in section \S\ref{irabs}.  
The fitting of isochrones to the CMD of NGC 752, resulting in our best 
estimate of the cluster age, is discussed in \S\ref{age}.  
We compare stellar model predictions to the observed cluster RC stars 
in \S\ref{isoch}, and summarize the main results of this investigation 
in \S\ref{cocl}.

\section{Membership Assignment Using \textit{Gaia} DR2}\label{gaia}

\begin{table*}                                                                                                   
 \begin{minipage}{170mm}                                      
 \caption{Kinematics and radial velocities.}
 \label{tab-motions}                                            
 \begin{tabular}{@{}lccccccc@{}}                             
 \hline                                                       
Star     &  Gaia DR2&  $\pi_{\rm (\it Gaia)}$  & $\mu_\alpha$ $_{\rm (\it Gaia)}$ & $\mu_\delta$ $_{\rm 
(\it Gaia)}$ & RV$^{\rm a}$  & RV$_{\rm (Paper~1)}$   &   RV$_{\rm (\it Gaia)}$  \\
& identifications  & (mas yr$^{-1}$) &  (mas yr$^{-1}$)   &  (mas yr$^{-1}$)   &
  (km s$^{-1}$) &  (km s$^{-1}$) & (km s$^{-1}$)               \\
 \hline
MMU 1	&	342554191959774720 & 2.081	$\pm$	0.049	&	9.780	$\pm$	0.082	&	$-12.003$	$\pm$	0.090	&	5.34	$\pm$ 0.24  &	4.73	$\pm$	0.20	&	5.29	$\pm$	0.35	\\
MMU 3	&	342554187663431424 &2.101	$\pm$	0.047	&	9.670	$\pm$	0.078	&	$-11.821$	$\pm$	0.082	&	4.69 $\pm$ 0.29  &	4.11	$\pm$	0.20	&	5.38	$\pm$	0.20	\\
MMU 11& 343118619382072832 &	2.251	$\pm$	0.066	&	9.811	$\pm$	0.094	&	$-12.244$	$\pm$	0.099	&	5.08 $\pm$ 0.26  &	4.45	$\pm$	0.19	&	5.63	$\pm$	0.11	\\
MMU 24 & 342929297223676160 &	2.158	$\pm$	0.049	&	9.602	$\pm$	0.102	&	$-11.940$	$\pm$	0.094	&	4.79 $\pm$ 0.18  &	4.86	$\pm$	0.19	&	5.70	$\pm$	0.23	\\
MMU 27 & 342536702852966784 &	2.174	$\pm$	0.049	&	9.620	$\pm$	0.084	&	$-11.701$	$\pm$	0.097	&	4.06 $\pm$ 0.18  &	4.39	$\pm$	0.19	&	4.93	$\pm$	0.18	\\
MMU 77 & 342532923281905408 &	2.205	$\pm$	0.050	&	9.827	$\pm$	0.077	&	$-11.905$	$\pm$	0.080	&	4.89 	$\pm$ 0.24  &	4.58	$\pm$	0.20	&	4.83	$\pm$	1.23	\\
MMU 137	& 342937195667536512 & 2.149      $\pm$	0.042	&	9.535	$\pm$	0.085	&	$-11.883$	$\pm$	0.082	&	5.17 $\pm$ 0.24  &	5.59	$\pm$	0.20	&	5.60	$\pm$	0.18	\\
MMU 295	& 342893803614055168 & 2.201      $\pm$	0.046	&	9.404	$\pm$	0.104	&	$-11.593$	$\pm$	0.087	&	5.29 $\pm$ 0.27  &	6.32	$\pm$	0.23	&				\\
MMU 311	& 342890127122193280 & 2.229      $\pm$	0.056	&	9.701	$\pm$	0.112	&	$-11.331$	$\pm$	0.106	&	5.69 $\pm$ 0.24  &	5.19	$\pm$	0.19	&	5.67	$\pm$	0.12	\\
MMU 1367 & 342899537393760512 & 2.251    $\pm$	0.052	&	9.688	$\pm$	0.094	&	$-11.794$	$\pm$	0.106	&	4.69 $\pm$ 0.24  &	3.98	$\pm$	0.19	&	5.09	$\pm$	0.23	\\
 \hline
RV (cluster mean) &  &     	&		&		&	4.97 $\pm$ 0.24  &	4.82	$\pm$	0.20	&	5.35	$\pm$	0.31	\\
 \hline
\multicolumn{5}{l}{$^{\rm a}$ This study.} \\
\end{tabular}
\end{minipage}                                                
\end{table*}

\begin{figure}
 \includegraphics[width=1\linewidth]{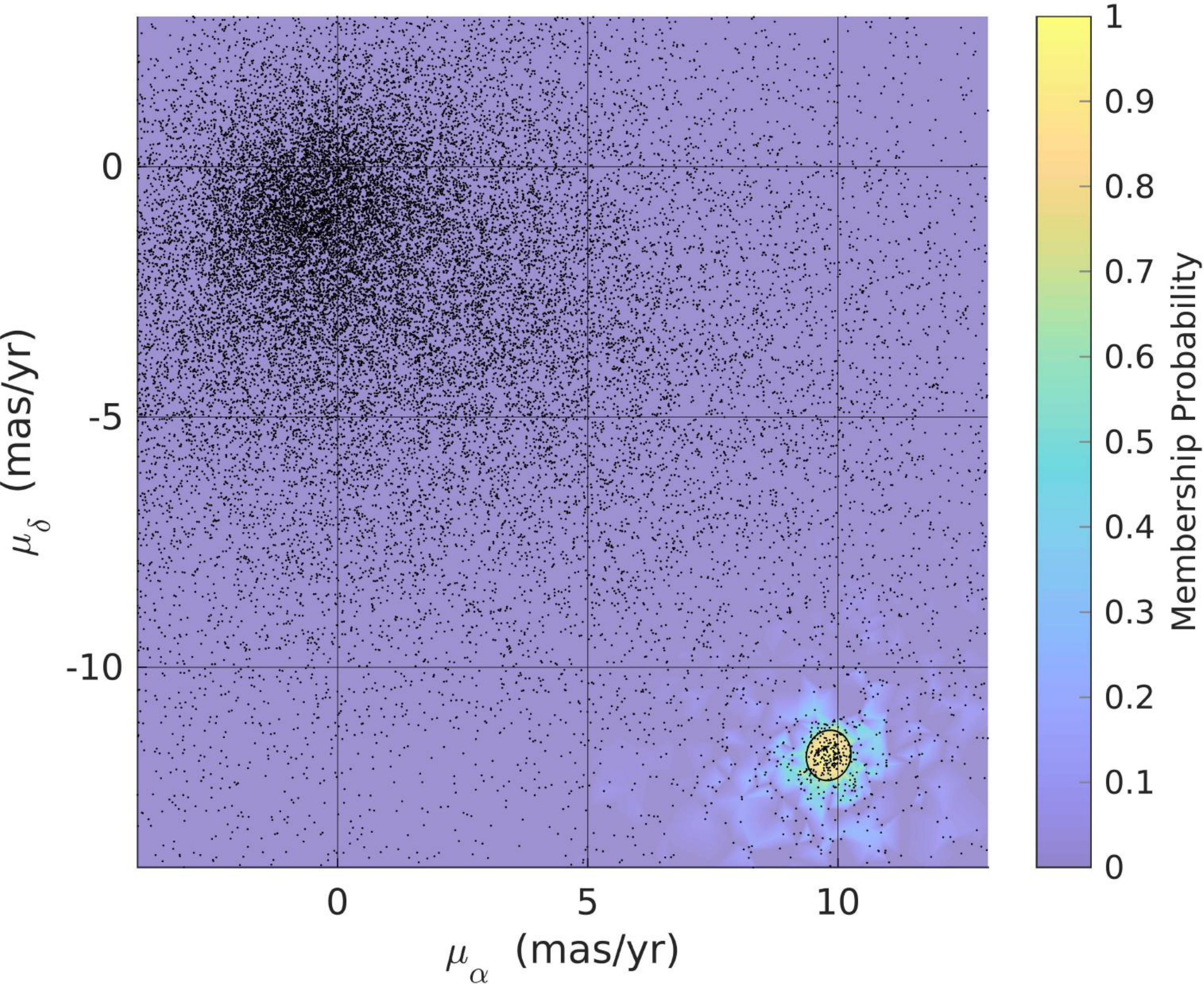}
      \caption{Vector point diagram of our data set, where 
               each dot shows the proper motion components of a star in right 
               ascension ($\mu_{\alpha}$) and declination ($\mu_{\delta}$). 
               The 95\% confidence ellipse in the lower right corner 
               encircles most of the cluster stars and represents the 
               intrinsic cluster center and dispersion. 
               The colors, defined in the side bar, signify NGC~752 membership 
               probabilities for each star.
               The total data set consisted of more stars more widely
               separated in proper motion from the field and cluster
               groups; they have been omitted for clarity in this plot.}
\label{membership_vpd}
\end{figure}

\begin{table*}
\begin{minipage}{165mm} 
 \caption{Basic parameters of the program stars and summary of IGRINS observations.} 
 \label{tab-basic}
 \begin{tabular}{@{}lccccccccccc@{}}
 \hline
Star     &  RA$^{\rm a}$   &  DEC$^{\rm a}$  &   V$^{\rm b}$   &   H$^{\rm c}$   &   K$^{\rm c}$   &
 G$^{\rm a}$  &  (G$_{\rm BP}-G_{\rm RP})^{\rm d}$ &  $(B - V)_{0}$ & $(V - K)_{0}$ &
 Date  &  S/N \\
      &  ($2000$) &  ($2000$)   &    &    &   & &     &   & &(UT) &  (s)      \\         
 \hline
MMU 1	&	01 55 12.62	&	37 50 14.55	&	9.50	&	7.37	&	7.23	&	9.23	&	1.13	&	0.92	&	2.17	&	02 12 2015	&	108	\\
MMU 3	&	01 55 15.29	&	37 50 31.30	&	9.57	&	7.32	&	7.20	&	9.28	&	1.16	&	0.96	&	2.27	&	02 12 2015	&	106	\\
MMU 11	&	01 55 27.67	&	37 59 55.24	&	9.29	&	7.16	&	7.04	&	9.03	&	1.12	&	0.93	&	2.15	&	02 12 2015	&	120	\\
MMU 24	&      01 55 39.37	&	37 52 52.51	&	8.92	&	6.67	&	6.55	&	8.65	&	1.16	&	0.98	&	2.28	&	03 12 2015	&	109	\\
MMU 27	&	01 55 42.39	&	37 37 54.57	&	9.16	&	6.90	&	6.80	&	8.88	&	1.17	&	0.98	&	2.27	&	03 12 2015	&	117	\\
MMU 77	&	01 56 21.64	&	37 36 08.43	&	9.38	&	7.05	&	6.92	&	9.09	&	1.21	&	0.99	&	2.36	&	03 12 2015	&	117	\\
MMU 137	&	01 57 03.11	&	38 08 02.65	&	8.93	&	6.66	&	6.54	&	8.64	&	1.18	&	0.99	&	2.29	&	04 12 2015	&	119	\\
MMU 295	&	01 58 29.82	&	37 51 37.57	&	9.30	&	7.17	&	7.04	&	9.05	&	1.13	&	0.93	&	2.17	&	04 12 2015	&	141	\\
MMU 311	&	01 58 52.90	&	37 48 57.23	&	9.06	&	6.80	&	6.64	&	8.77	&	1.20	&	1.00	&	2.33	&	04 12 2015	&	113	\\
MMU 1367&	01 59 14.80	&	38 00 55.29	&	9.01	&	6.77	&	6.65	&	8.72	&	1.18	&	0.98	&	2.26	&	26 11 2018	&	113	\\
 \hline
 \multicolumn{9}{l}{$^{\rm a}$ \cite{GAIA18b}}\\
\multicolumn{9}{l}{$^{\rm b}$ \cite{dan94}} \\
\multicolumn{9}{l}{$^{\rm c}$ \cite{2MASS}}\\
\multicolumn{9}{l}{$^{\rm d}$ \cite{cantat18}}\\
\end{tabular}
\end{minipage}
\end{table*}

For determination of our NGC~752 member set, we created 
a Gaussian mixture model. 
This model was built using proper motion data from the \textit{Gaia} \citep{GAIA16} 
Data Release 2 \citep{GAIA18b}. 
All stars that had \textit{Gaia} DR2 proper motions and resided within 
75$^\prime$ of the approximate NGC~752 cluster center, 
$\alpha_{2000} = 1^h57^m41.0^s$ and 
$\delta_{2000} = +37^\circ47^\prime6^{\prime\prime}$, 
were considered for membership. 
A vector point diagram for these stars is shown in Figure \ref{membership_vpd}. 
Each dot represents the proper motion components of a single candidate star.

Fitting mixture models is an applied statistical method 
that allows for the 
Bayesian determination of membership probabilities for individual stars. 
Open cluster applications of membership models like ours date at least back to 
\cite{sanders71}, whose model is fundamentally similar to ours: the sum of 
two normal probability densities $-$ bivariate in the case of proper motion data
$-$ is fit to observed right ascension and declination proper motion components.
Our probability density function is of the form
\begin{align}
\label{eq:model_form}
    \Phi(\mu_{x_i}, \mu_{y_i}, \epsilon_{x_i}, \epsilon_{y_i}) = \phi_c + \phi_f \textnormal{,}
\end{align}

\noindent where $\mu_{x_i}$ and $\mu_{y_i}$ are the 
proper motion components 
for the $i^{th}$ star in our data set, and $\epsilon_{x_i}$ and 
$\epsilon_{y_i}$ are their respective errors.

$\phi_c$ and $\phi_f$ are Gaussians and represent the 
cluster and field star distributions, respectively. 
Both Gaussians are symmetrical and elliptical. 
It is common to make the assumption of a circular cluster distribution, 
but we have left it as elliptical for increased accuracy. 
In addition, we have adopted a method derived by \cite{zhao90} that takes 
into account not only the intrinsic dispersions of $\phi_c$ and $\phi_f$ 
but also the observed \textit{Gaia} DR2 errors for each individual star. 
In total, there are 11 parameters needed to characterize the 
distributions $\phi_c$ and $\phi_f$ in our model. 
After we solved for these, cluster membership probabilities were calculated. 
Figure \ref{membership_vpd} presents the final probabilities for stars 
within the proper motion ranges plotted. 
See the appendix for more detail on the exact forms of $\phi_c$ and 
$\phi_f$ and the techniques used to solve for the parameters.

In a mixture model such as ours, it is inevitable 
that some field stars will 
pass the cluster membership probability cutoff simply due to random chance. 
To mitigate this effect, we imposed parallax bounds on the remaining stars. 
By examination of the parallax density of the stars remaining after the proper 
motion cutoff, we found that the parallax distribution for NGC~752 peaked 
at $\simeq$2.235 mas with a base width of $\simeq$0.8 mas. 
We therefore set parallax bounds of 1.735 mas to 2.735 mas. 
All stars that passed a 50\% proper motion model cutoff and fell within these 
parallax bounds were considered to be physical cluster members.

Our membership study was performed independently 
of any other NGC~752 membership analysis. 
Our computations yield cluster parameters of
$\mu_\alpha$ = 9.827$\pm$0.017~mas/yr, 
$\mu_\delta$ = $-$11.782$\pm$0.019~mas/yr, and
parallax  $p$= 2.229$\pm$0.009~mas.
This corresponds to a distance of
448 pc and a true distance modulus $(m-M)_0 = 8.26$, which has been
adopted in the  fitting of isochrones to the observed CMD in \S\ref{age}.
We caution the reader that our parallax uncertainty for 
NGC~752 is purely statistical;  a more realistic estimate would take 
into account possible systematic \textit{Gaia} uncertainties, which \cite{arenou18} 
suggest can be as large as 0.03~mas.
Recently \cite{cantat18} have performed a membership study of NGC~752 using 
very different methods. 
They suggest cluster parameters
$\mu_\alpha$ = 9.810$\pm$0.019~mas/yr, 
$\mu_\delta$ = $-$11.713$\pm$0.019~mas/yr, and 
$p$ = 2.239$\pm$0.005~mas. 
Although our study focused on proper motions and we do not claim to have 
determined precise parallax estimates, our results are 
essentially in agreement with their work.

In Table~\ref{tab-motions} we have listed the program stars with their
\textit{Gaia} DR2 identifications, parallaxes and proper motions. 
Table~\ref{tab-motions} also contains radial velocities (RVs) from 
\textit{Gaia} \citep{GAIA18b}, optical (BT15) and $IR$ spectra. 
The $IR$ RVs were measured applying a similar method described in Paper~1 
using at least 10 spectral orders that are less affected by the atmospheric 
telluric lines.
The mean cluster RVs from these measurements
agree well within the mutual uncertainties:
$\langle RV\rangle_{\rm opt}$ $=$ $4.82 \pm 0.20$ kms$^{-1}$ $(\sigma=0.71)$,
$\langle RV\rangle_{\rm IR}$ $=$ $4.97 \pm 0.24$ kms$^{-1}$ $(\sigma=0.45)$, 
and 
$\langle RV\rangle_{\rm Gaia}$ $=$ $5.35 \pm 0.31$ kms$^{-1}$ $(\sigma=0.33)$
(from 9 RGs).

\section{Observations and Data Reduction}\label{obs}

\begin{table*}
 \begin{minipage}{173mm}
 \caption{Model atmosphere parameters and [Fe/H] abundances from both wavelength regions.} 
 \label{tab-model}
  \begin{tabular}{@{}lcccccrccrccrcc@{}}
  \hline
 Star     &  \teff\ & \teff$_{(\rm \it Gaia)}$ &  \teff$_{\rm (LDR)}$ & \logg &     $\xi_{t}$ & 
     [\ion{Fe}{i}/H]  & $\sigma$ & \# & 
     [\ion{Fe}{ii}/H] & $\sigma$ & \# &
     [\ion{Fe}{i}/H]  & $\sigma$ & \# \\
          &    (K)  &        (K)  &    (K) &   & (km s$^{-1}$) &
                    (opt) &    (opt) &   (opt) &
                    (opt) &    (opt) &   (opt) &
                     ($IR$) &     ($IR$) &    ($IR$) \\
 \hline
MMU 1	&	5005	&	4929	&	5075	&	2.95	&	1.07	&$	0.04	$&	0.07	&	58	&$	-0.02	$&	0.03	&	12	&	0.00	&	0.06	&	17	\\
MMU 3	&	4886	&	4953	&	5010	&	2.76	&	1.10	&$	-0.05	$&	0.07	&	62	&$	-0.07	$&	0.06	&	10	&    -0.03	&	0.05	&	20	\\
MMU 11	&	4988	&	4956	&	5045	&	2.80	&	1.14	&$	0.03	$&	0.07	&	59	&$	-0.01	$&	0.04	&	11	&    -0.01	&	0.06	&	21	\\
MMU 24	&	4839	&	4914	&	4986	&	2.42	&	1.23	&$	-0.05	$&	0.07	&	57	&$	-0.11	$&	0.04	&	12	&    -0.05	&	0.05	&	20	\\
MMU 27	&	4966	&	4878	&	4948	&	2.73	&	1.16	&$	0.08	$&	0.07	&	57	&$	-0.06	$&	0.03	&	11	&     0.06	&	0.05	&	20	\\
MMU 77	&	4874	&	4850	&	4944	&	2.80	&	1.15	&$	0.04	$&	0.08	&	62	&$	-0.06	$&	0.06	&	11	&     0.03	&	0.04	&	19	\\
MMU 137	&	4832	&	4850	&	4970	&	2.51	&	1.29	&$	-0.08	$&	0.06	&	58	&$	-0.16	$&	0.05	&	9	&    -0.09	&	0.04	&	19	\\
MMU 295	&	5039	&	5050	&	5053	&	2.88	&	1.10	&$	0.07	$&	0.06	&	58	&$	-0.01	$&	0.04	&	10	&     0.03	&	0.06	&	20	\\
MMU 311	&	4874	&	4846	&	4959	&	2.68	&	1.24	&$	0.07	$&	0.09	&	62	&$	0.01	$&	0.06	&	10	&     0.05	&	0.06	&	21	\\
MMU 1367&	4831	&	4831	&	4985	&	2.42	&	1.22	&$	-0.02	$&	0.07	&	59	&$	-0.08	$&	0.03	&	12	&    -0.02	&	0.08	&	19	\\
 \hline
\end{tabular}
\end{minipage}
\end{table*}

We gathered IGRINS $H$- and $K$-band high resolution 
spectra for the 10 NGC~752 RG members studied in the optical spectral region by BT15.
The stars chosen for that paper were selected from the radial velocity membership catalog 
of \cite{mermilliod08}, before the release of Gaia DR2 astrometric data. The membership 
analysis presented here confirms that our targets belong to NGC~752.  
Additionally, several stars not included here appear to be RG members 
(this study and B. Twarog, private communication). 
Future spectroscopic study of these stars would be welcome.
The log of the IGRINS observations is given in Table~\ref{tab-basic} along 
with the basic parameters of program stars. 
These stars are all red giants with similar parameters, as indicated by
spectroscopic analyses (\teff~$\sim$~4900~K, \logg~$\sim$~2.7; BT15) and 
by photometric data ($V$~$\simeq$~9.2, $M_V$~$\simeq$~1.0, 
$(B-V)_0$~$\simeq$~0.97).
Three other stars with similar photometric characteristics satisfy our 
NGC~752 membership criteria: BD+37~404 (MMU~2054), BD+36 328 (MMU~1533), and 
BD+37 422 (MMU 110, HD 11811).
The derived distance for BD+37 422 is $\sim$30~pc from the cluster mean,
but this star is a known spectroscopic binary, so its photometric and
astrometric data should be treated with caution.
None of these three stars appears to have been subjected to a comprehensive 
atmospheric and abundance analysis; future spectroscopic studies of them
would be of some interest.

Characteristics of the IGRINS instrument have been presented 
by \cite{yuk10} and \cite{park14}.
This spectrograph employs a silicon immersion grating (\citealt{gully12}) 
to achieve resolving power $R$~$\equiv$~$\lambda/\Delta\lambda$ $\simeq$ 45000
for the entire $H$ and $K$ bands (1.45 -- 2.5 $\micron$) in a single exposure. 
Almost all of the observations were made with IGRINS installed at the 
Cassegrain focus of the 2.7m Harlan J. Smith Telescope at McDonald 
Observatory in 2015 December (\citealt{mace16}). 
One object, MMU 1367, was observed with IGRINS on Lowell Observatory's 
4.3m Discovery Channel Telescope (\citealt{mace18}). 
Typical exposure times were 300s and 
used ABBA nod sequences along the spectrograph slit length.
We also observed telluric standards with spectral types of B9IV to A0V.
They were observed right after each science exposures at very close 
airmasses to the ones at which program stars were observed. 
The spectra used in this analysis were reduced using the IGRINS 
pipeline (\citealt{lee17}). 
The pipeline performs flat-field correction, A-B frame subtractions to 
remove skyline emission, wavelength correction using OH emission and 
telluric absorption, and optimal spectral extraction. 
Due to their high rotational velocities ($\sim$150 \kmsec), 
telluric stars come with extremely broadened absorption features that can be easily 
distinguished from the atmospheric telluric lines.
After removing the extremely broadened features from the spectra of 
telluric standards, we used the \textit{telluric} task of IRAF\footnote{
\url{http://iraf.noao.edu/}}
to remove the contamination of atmospheric absorption lines from
the spectra of our program stars.

\section{Model Atmospheres and Abundances from the Optical Region}\label{modopt}

Model atmospheric parameters (Table~\ref{tab-model}) of the 10 RG members 
of NGC 752 were previously presented in BT15, along with the abundances 
for 26 species of 23 elements present in the optical spectral region
(see Table 10 in BT15). 
In this study, we newly report abundances of species 
\ion{S}{i}, \ion{K}{i}, and \ion{Ce}{ii} from those spectra. 
We present optical sulfur abundances for our targets for the 
first time in NGC\,752 using the \ion{S}{i} triplet centered at 6757.17 \AA. 
\cite{takeda16} reported non-local thermodynamic equilibrium (non-LTE) 
corrections for \species{S}{i}, estimated them to be $\lesssim 0.1$ dex for 
G-K giants for this blended S feature.
We have also repeated the analyses for optical \ion{Sc}{ii} lines, 
adopting new \loggf\ values from \cite{lawler19}.
Detailed description of the elemental abundance analysis methods in the 
optical region were provided in BT15, in which we also derived the solar 
photospheric abundances following the same procedure applied for the 
program stars, to obtain the differential values of stellar abundances 
relative to the Sun.
Here we used the same method as in Paper~1, adopting the \cite{asplund:09}
solar photospheric abundances for both regions in order to achieve 
consistency between optical and $IR$ data sets. 
These slightly revised relative optical abundances 
are listed in the upper part of Table~\ref{tab-abunds}.
Mean abundances of the species and their standard deviations are 
given in columns 12 and 13 of this table.
For the [X/Fe] calculations, we took star-by-star species differences 
using both \ion{Fe}{i} and \ion{Fe}{ii} abundances as appropriate.
We will discuss the differences between the NGC\,752 and NGC\,6940 abundance
sets in \S\ref{comp6940}.

\begin{figure}
 \includegraphics[width=\columnwidth]{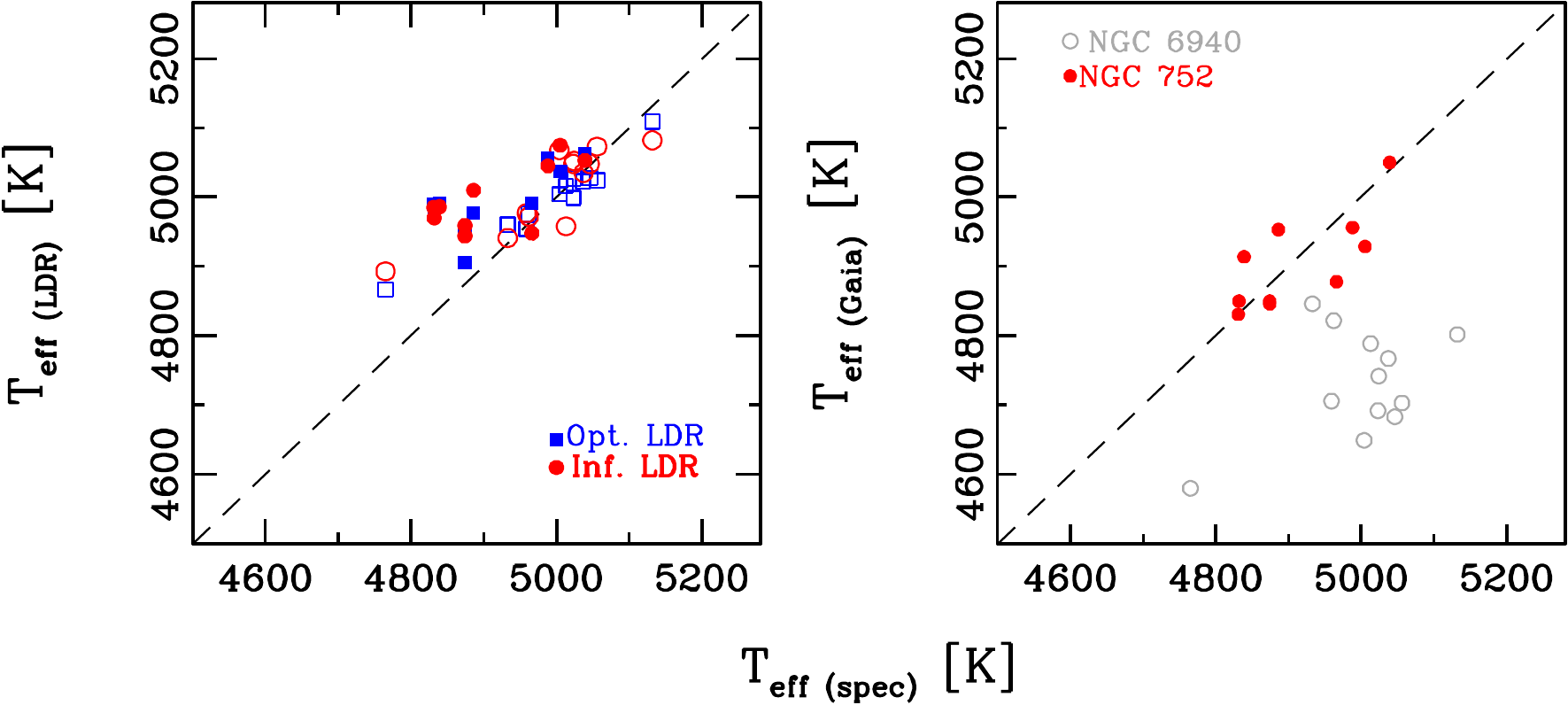}
      \caption{Left panel: comparison of optical and $H$-band LDR \teff\ values 
               with those derived spectroscopically in Paper~I. 
               Full symbols represents NGC~752 RGs, open symbols are NGC~6940 
               RGs.
               Circle's (red) are IR LDR and square symbols (blue) are optical 
               LDR \teff\ values.  
               Right panel: comparison of \textit{Gaia} temperatures along with the 
               spectroscopic \teff's of the NGC~752 (full red circles) 
               and NGC~6940 (open grey symbols). 
               The dashed line represents equality of the temperatures 
               for both panels.}
      \label{teff}
\end{figure}

\section{Temperature determination using IGRINS Data}\label{ldrcomp}

Accurate effective temperatures, gravities, and microturbulent velocities
are required for abundance analyses. 
In our OC studies we have adopted traditional line-by-line equivalent width 
($EW$) analyses to derive atmospheric parameters \teff, \logg, \vmicro, and
[M/H].
For NGC\,752 these parameters derived in BT15 are listed in 
Table~\ref{tab-model}
and we have used them for all of the abundances reported in this paper.
However, for heavily dust-obscured clusters optical parameter determinations 
will not be possible, and $IR$-based methods will be needed.
Here we explore $IR$ \teff\ estimates.

Line-depth ratios (LDR) have proven to be good temperature indicators in 
several studies (e.g. \citealt{gray91,kovt06,biazzo07a,biazzo07b}).
The LDR method is based on depth ratios of high-excitation 
atomic lines (relatively sensitive to \teff) to low-excitation lines 
(much less sensitive to \teff).
This spectroscopic method has some attractive features: LDR temperatures 
are not affected by interstellar reddening and extinction, and they also only 
weakly depend on other atmospheric parameters for solar metallicity RGs. 
In our previous optical studies (BT15, BT16), we used the line
pairs and equations of \cite{biazzo07a,biazzo07b}. 
Recently, \cite{fukue15} have applied the LDR method to the $IR$ spectra,
and have found nine pairs of absorption lines in the $H$-band 
(1.4 $-$ 1.8 $\mu$m) to be good \teff\ indicators. 
The application of LDR method to the $IR$ spectra brings new opportunities, 
such as access to the most dust obscured stars and the determination of the 
\teff~without any information from the optical region. 

We applied the \cite{fukue15} relationships to NGC\,6940 IGRINS data
in Paper~1, and now we have calculated the LDR effective temperatures also 
for the NGC~752 RGs.
\teff\ results from this LDR method are listed in Table~\ref{tab-model} 
along with other \teff~values obtained from the optical region. 
In Figure~\ref{teff} (left panel), we compare the temperatures derived from 
both optical and $IR$ LDRs with the spectroscopic temperatures. 
This comparison indicates that $T_{\rm eff,spec}$ values derived from 
traditional line-by-line Fe and Ti $EW$ analyses and 
$T_{\rm eff,LDR}$ agree well for \teff~$\geq$~4900~K.
For cooler RG stars the LDR temperatures become systematically larger.
The $T_{\rm eff,LDR}$ of MMU\,1367, the coolest ($T_{\rm eff,spec}$ = 4831 K) 
member among others, is 154 K away from its spectroscopic temperature.
This star and MMU\,311 also deviate similarly in the optical (BT15, Table~6).
The $IR$-based LDR temperatures of NGC\,6940 shown in the left panel of 
Figure~\ref{teff} suggest a similar effect in that cluster also.

The LDR calibration issue is not of importance in our work, as most program
stars are warmer than 4900~K, but it should be revisited in the
future with larger sets of spectroscopic data in the $IR$.
Considering the NGC\,752 sample as a whole, on average LDR and spectroscopic 
temperature are in reasonable accord:
$\langle T_{\rm eff,LDR} - T_{\rm eff,spec}\rangle_{\rm IR}$ $=$ $84 \pm 18$ K, 
and 
$\langle T_{\rm eff,LDR} - T_{\rm eff,spec}\rangle_{\rm opt}$ $=$ $81 \pm 18$ K.
Overall the $IR$ LDRs provide reliable temperatures in the absence of 
information from the optical region for giant stars with solar metallicities 
for the temperature range considered here.

The right panel of Figure~\ref{teff} shows comparisons of \textit{Gaia} temperatures 
(Table~\ref{tab-model}) vs. $T_{\rm eff,spec}$ for the RGs of NGC\,752 
and NGC\,6940.
By inspection, \textit{Gaia} and spectroscopic agree well for NGC\,752, and for the
whole sample 
$\langle T_{\rm eff, Gaia} - T_{\rm eff, spec.}\rangle_{\rm NGC~752}$ $=$ $-8 \pm 17$~K. 
For NGC\,6940 the lack of agreement between spectroscopic and \textit{Gaia} 
temperatures is clear in Figure~\ref{teff}:
$\langle T_{\rm eff, Gaia} - T_{\rm eff, spec.}\rangle_{\rm NGC~6940}$ $=$ $-264 \pm 26$ K.
\textit{Gaia} temperatures are photometrically based and thus depend on interstellar 
extinction corrections.
For NGC\,6940 $E(B-V)$ = 0.21, while for NGC\,752 the reddening is very small,
$E(B-V)$ = 0.035 .
This probably is related to the poor \teff\ correlation for NGC\,6940, but
resolution of the question is beyond the scope of this paper.

\begin{table*}
  \caption{Relative abundances and \ciso\ ratios of NGC 752 RGs and their cluster means.}
   \label{tab-abunds}
  \begin{tabular}{@{}lrrrrrrrrrrrcl@{}}
  \hline
Species  & \multicolumn{10}{c}{MMU} \\
$\rm{[X/Fe]}$& 1  &      3  &      11  &      24  &      27  &     77  &     137  &     295  &     311  &     1367   &        mean & $\sigma$ & \#$_{max}$ \\ 
\hline
\multicolumn{14}{c}{Optical Spectral Region} \\
\ion{Na}{i}	&$	0.20	$&$	0.16	$&$	0.21	$&$	0.27	$&$	0.16	$&$	0.15	$&$	0.30	$&$	0.17	$&$	0.20	$&$	0.23	$&$	0.20	$&	0.05	&	4	\\
\ion{Mg}{i}	&$	-0.06	$&$	0.03	$&$	0.00	$&$	0.08	$&$	-0.02	$&$	-0.02	$&$	0.03	$&$	-0.09	$&$	-0.07	$&$	0.00	$&$	-0.01	$&	0.05	&	2	\\
\ion{Al}{i}	&$	-0.08	$&$	-0.02	$&$	-0.08	$&$	0.00	$&$	-0.08	$&$	-0.04	$&$	0.02	$&$	-0.09	$&$	-0.07	$&$	-0.03	$&$	-0.04	$&	0.04	&	2	\\
\ion{Si}{i}	&$	0.18	$&$	0.22	$&$	0.20	$&$	0.27	$&$	0.17	$&$	0.22	$&$	0.29	$&$	0.19	$&$	0.22	$&$	0.27	$&$	0.22	$&	0.04	&	15	\\
\ion{S}{i}$^{*}$	&$	0.02	$&$	0.13	$&$	0.04	$&$	0.04	$&$	-0.01	$&$	0.07	$&$	0.17	$&$	-0.04	$&$	0.04	$&$	0.06	$&$	0.05	$&	0.06	&	2	\\
\ion{K}{i}$^{*}$	&$	0.47	$&$	0.61	$&$	0.49	$&$	0.54	$&$	0.51	$&$	0.44	$&$	0.47	$&$	0.51	$&$	0.47	$&$	0.52	$&$	0.50	$&	0.05	&	1	\\
\ion{Ca}{i}	&$	0.13	$&$	0.17	$&$	0.12	$&$	0.16	$&$	0.09	$&$	0.13	$&$	0.13	$&$	0.11	$&$	0.12	$&$	0.15	$&$	0.13	$&	0.02	&	10	\\
\ion{Sc}{ii}$^{*}$	&$	0.08	$&$	0.02	$&$	0.06	$&$	0.06	$&$	0.14	$&$	0.12	$&$	0.09	$&$	0.08	$&$	0.09	$&$	0.03	$&$	0.08	$&	0.01	&	6	\\
\ion{Ti}{i}	&$	-0.04	$&$	-0.05	$&$	-0.08	$&$	-0.09	$&$	-0.01	$&$	-0.01	$&$	-0.07	$&$	-0.02	$&$	-0.06	$&$	-0.11	$&$	-0.05	$&	0.03	&	11	\\
\ion{Ti}{ii}	&$	0.14	$&$	0.07	$&$	0.10	$&$	0.03	$&$	0.11	$&$	0.04	$&$	0.02	$&$	0.13	$&$	0.02	$&$	0.01	$&$	0.07	$&	0.05	&	4	\\
\ion{V}{i}	&$	-0.03	$&$	-0.04	$&$	-0.08	$&$	-0.08	$&$	-0.02	$&$	0.01	$&$	-0.06	$&$	-0.04	$&$	-0.04	$&$	-0.11	$&$	-0.05	$&	0.04	&	12	\\
\ion{Cr}{i}	&$	0.07	$&$	0.10	$&$	0.03	$&$	0.07	$&$	0.07	$&$	0.05	$&$	0.09	$&$	-0.01	$&$	-0.01	$&$	0.04	$&$	0.05	$&	0.04	&	14	\\
\ion{Cr}{ii}	&$	0.17	$&$	0.17	$&$	0.18	$&$	0.26	$&$	0.20	$&$	0.23	$&$	0.20	$&$	0.25	$&$	0.19	$&$	0.18	$&$	0.20	$&	0.03	&	3	\\
\ion{Mn}{i}	&$	-0.17	$&$	-0.23	$&$	-0.18	$&$	-0.15	$&$	-0.24	$&$	-0.21	$&$	-0.22	$&$	-0.30	$&$	-0.24	$&$	-0.07	$&$	-0.20	$&	0.06	&	3	\\
\ion{Co}{i}	&$	-0.07	$&$	-0.07	$&$	-0.10	$&$	-0.08	$&$	-0.08	$&$	-0.05	$&$	-0.06	$&$	-0.10	$&$	-0.06	$&$	-0.09	$&$	-0.07	$&	0.02	&	5	\\
\ion{Ni}{i}	&$	0.09	$&$	0.07	$&$	0.07	$&$	0.09	$&$	0.10	$&$	0.12	$&$	0.09	$&$	0.12	$&$	0.11	$&$	0.07	$&$	0.09	$&	0.02	&	29	\\
\ion{Cu}{i}	&$	-0.18	$&$	-0.31	$&$	-0.26	$&$	-0.26	$&$	-0.27	$&$	-0.17	$&$	-0.28	$&$	-0.37	$&$	-0.31	$&$	-0.30	$&$	-0.27	$&	0.06	&	1	\\
\ion{Zn}{i}	&$	-0.07	$&$	0.02	$&$	-0.02	$&$	0.12	$&$	-0.04	$&$	0.03	$&$	0.09	$&$	-0.05	$&$	0.05	$&$	0.04	$&$	0.02	$&	0.06	&	1	\\
\ion{Y}{ii}	&$	-0.02	$&$	-0.04	$&$	0.05	$&$	0.08	$&$	0.04	$&$	0.10	$&$	-0.05	$&$	0.00	$&$	-0.04	$&$	-0.06	$&$	0.01	$&	0.06	&	4	\\
\ion{La}{ii}	&$	0.22	$&$	0.17	$&$	0.17	$&$	0.21	$&$	0.25	$&$	0.36	$&$	0.26	$&$	0.20	$&$	0.19	$&$	0.18	$&$	0.22	$&	0.06	&	4	\\
\ion{Ce}{ii}$^{*}$	&$	0.08	$&$	0.10	$&$	0.07	$&$	0.09	$&$	0.18	$&$	0.14	$&$	0.14	$&$	0.08	$&$	0.10	$&$	0.08	$&$	0.11	$&	0.03	&	4	\\
\ion{Nd}{ii}	&$	0.20	$&$	0.04	$&$	0.24	$&$	0.22	$&$	0.33	$&$	0.33	$&$	0.27	$&$	0.19	$&$	0.20	$&$	0.16	$&$	0.22	$&	0.08	&	3	\\
\ion{Eu}{ii}	&$	0.11	$&$	0.21	$&$	0.05	$&$	0.21	$&$	0.16	$&$	0.25	$&$	0.21	$&$	0.10	$&$	0.15	$&$	0.14	$&$	0.16	$&	0.06	&	2	\\
log~$\epsilon$(Li)	&$	0.15	$&$	1.25	$&$	1.00	$&$	<0.0	$&$	0.95	$&$	1.34	$&$	<0.0	$&$	<0.0	$&$	0.78	$&$	<0.0	$&$		$&		&	1	\\
\ciso	&$	25	$&$	20	$&$	25	$&$	13	$&$	17	$&$	25	$&$	15	$&$	20	$&$	15	$&$	20	$&$     19.5    $&$     4.5     $&      CN	\\
C 	&$	-0.39	$&$	-0.28	$&$	-0.27	$&$	-0.27	$&$	-0.37	$&$	-0.39	$&$	-0.21	$&$	-0.32	$&$	-0.37	$&$	-0.28	$&$	-0.31	$&	0.06	&	C$_{2}$, CH, \ion{C}{i}	\\
N 	&$	0.51	$&$	0.45	$&$	0.47	$&$	0.50	$&$	0.50	$&$	0.46	$&$	0.48	$&$	0.48	$&$	0.48	$&$	0.47	$&$	0.48	$&	0.02	&	CN	\\
O 	&$	-0.15	$&$	-0.16	$&$	-0.14	$&$	-0.14	$&$	-0.11	$&$	-0.10	$&$	-0.10	$&$	-0.11	$&$	-0.13	$&$	-0.14	$&$	-0.13	$&	0.02	&	[O I]	\\
	&		&		&		&		&		&		&		&		&		&		&		&		&		\\
\multicolumn{14}{c}{IGRINS H \& K  Spectral Region} \\
\ion{Na}{i}&$	0.11	$&$	0.02	$&$	0.15	$&$	0.16	$&$	0.10	$&$	0.12	$&$	0.11	$&$	0.11	$&$	0.07	$&$	0.19	$&$	0.12	$&	0.05	&	5	\\
\ion{Mg}{i}&$	-0.05	$&$	-0.01	$&$	-0.02	$&$	0.01	$&$	-0.03	$&$	0.03	$&$	0.02	$&$	-0.05	$&$	-0.01	$&$	0.02	$&$	-0.01	$&	0.03	&	11	\\
\ion{Al}{i}	&$	0.02	$&$	-0.05	$&$	0.04	$&$	0.04	$&$	0.01	$&$	0.05	$&$	0.04	$&$	0.04	$&$	0.00	$&$	0.04	$&$	0.02	$&	0.03	&	6	\\
\ion{Si}{i}	&$	0.07	$&$	0.10	$&$	0.13	$&$	0.14	$&$	0.10	$&$	0.08	$&$	0.15	$&$	0.10	$&$	0.05	$&$	0.16	$&$	0.11	$&	0.04	&	11	\\
\ion{P}{i}	&$	-0.08	$&$	0.16	$&$	0.06	$&$	0.15	$&$	-0.11	$&$	0.09	$&$	0.09	$&$	0.04	$&$	0.07	$&$	-0.01	$&$	0.05	$&	0.09	&	2	\\
\ion{S}{i}	&$	0.02	$&$	0.04	$&$	0.05	$&$	0.03	$&$	-0.05	$&$	-0.02	$&$	0.10	$&$	0.02	$&$	-0.01	$&$	-0.01	$&$	0.02	$&	0.04	&	10	\\
\ion{K}{i}	&$	-0.03	$&$	0.02	$&$	-0.04	$&$	-0.07	$&$	-0.11	$&$	0.00	$&$	-0.11	$&$	-0.24	$&$	-0.08	$&$	0.00	$&$	-0.06	$&	0.08	&	2	\\
\ion{Ca}{i}	&$	0.12	$&$	0.10	$&$	0.10	$&$	0.11	$&$	0.07	$&$	0.13	$&$	0.10	$&$	0.05	$&$	0.06	$&$	0.14	$&$	0.10	$&	0.03	&	11	\\
\ion{Sc}{i}	&$	0.01	$&$	-0.11	$&$	-0.02	$&$	-0.14	$&$	-0.10	$&$	-0.02	$&$	0.01	$&$	-0.06	$&$	0.04	$&$	-0.05	$&$	-0.04	$&	0.06	&	2	\\
\ion{Ti}{i}	&$	-0.02	$&$	-0.09	$&$	-0.05	$&$	-0.09	$&$	-0.05	$&$	0.02	$&$	-0.06	$&$	-0.10	$&$	-0.07	$&$	-0.15	$&$	-0.06	$&	0.05	&	10	\\
\ion{Ti}{ii}	&$	-0.12	$&$	-0.08	$&$	-0.13	$&$	-0.17	$&$	-0.16	$&$	-0.10	$&$	-0.12	$&$	-0.16	$&$	-0.12	$&$	-0.22	$&$	-0.14	$&	0.04	&	1	\\
\ion{Cr}{i}	&$	0.00	$&$	-0.04	$&$	0.00	$&$	-0.03	$&$	-0.10	$&$	-0.10	$&$	0.00	$&$	-0.13	$&$	-0.03	$&$	-0.08	$&$	-0.05	$&	0.05	&	3	\\
\ion{Co}{i}&$	-0.01	$&$	0.11	$&$	-0.11	$&$	-0.06	$&$	-0.03	$&$	0.02	$&$	0.07	$&$	0.11	$&$	0.03	$&$	-0.09	$&$	0.01	$&	0.08	&	1	\\
\ion{Ni}{i}	&$	0.04	$&$	-0.02	$&$	0.03	$&$	0.03	$&$	0.00	$&$	0.03	$&$	-0.01	$&$	-0.03	$&$	0.01	$&$	-0.02	$&$	0.01	$&	0.03	&	6	\\
\ion{Ce}{ii}&$	0.13	$&$	0.05	$&$	0.13	$&$	0.10	$&$	0.06	$&$	0.20	$&$	0.15	$&$	0.16	$&$	0.13	$&$	0.02	$&$	0.11	$&	0.06	&	9	\\
\ion{Nd}{ii}&$	0.43	$&$	0.15	$&$	0.34	$&$		$&$	0.16	$&$	0.19	$&$	0.30	$&$	0.35	$&$	0.21	$&$	0.05	$&$	0.24	$&	0.12	&	1	\\
\ion{Yb}{ii}&$	0.10	$&$	0.04	$&$	0.06	$&$	0.02	$&$	-0.06	$&$	0.07	$&$	0.04	$&$	0.10	$&$	0.00	$&$	0.00	$&$	0.04	$&	0.05	&	1	\\
\ciso&$	28	$&$	28	$&$	28	$&$	20	$&$	22	$&$	30	$&$	20	$&$	25	$&$	23	$&$	16	$&$     25.0    $&      3.6&	CO	\\
C	&$	-0.32	$&$	-0.41	$&$	-0.31	$&$	-0.37	$&$	-0.32	$&$	-0.30	$&$	-0.27	$&$	-0.31	$&$	-0.31	$&$	-0.36	$&$	-0.33	$&	0.04	&	CO, \ion{C}{i}	\\
N	&$	0.58	$&$	0.48	$&$	0.49	$&$	0.45	$&$	0.41	$&$	0.37	$&$	0.42	$&$	0.48	$&$	0.39	$&$	0.31	$&$	0.44	$&	0.08	&	CN	\\
O	&$	0.13	$&$	0.05	$&$	0.01	$&$	-0.06	$&$	0.06	$&$	0.05	$&$	-0.04	$&$	0.13	$&$	0.00	$&$	-0.14	$&$	0.02	$&	0.08	&	OH	\\
\hline                                                                                                                                                          
\multicolumn{14}{l}{$^{*}$ This study.}
\end{tabular}                                                                                                                                                   
\end{table*}

\section{Abundances from the Infrared Region}\label{irabs}

\begin{figure}
 \includegraphics[width=\columnwidth]{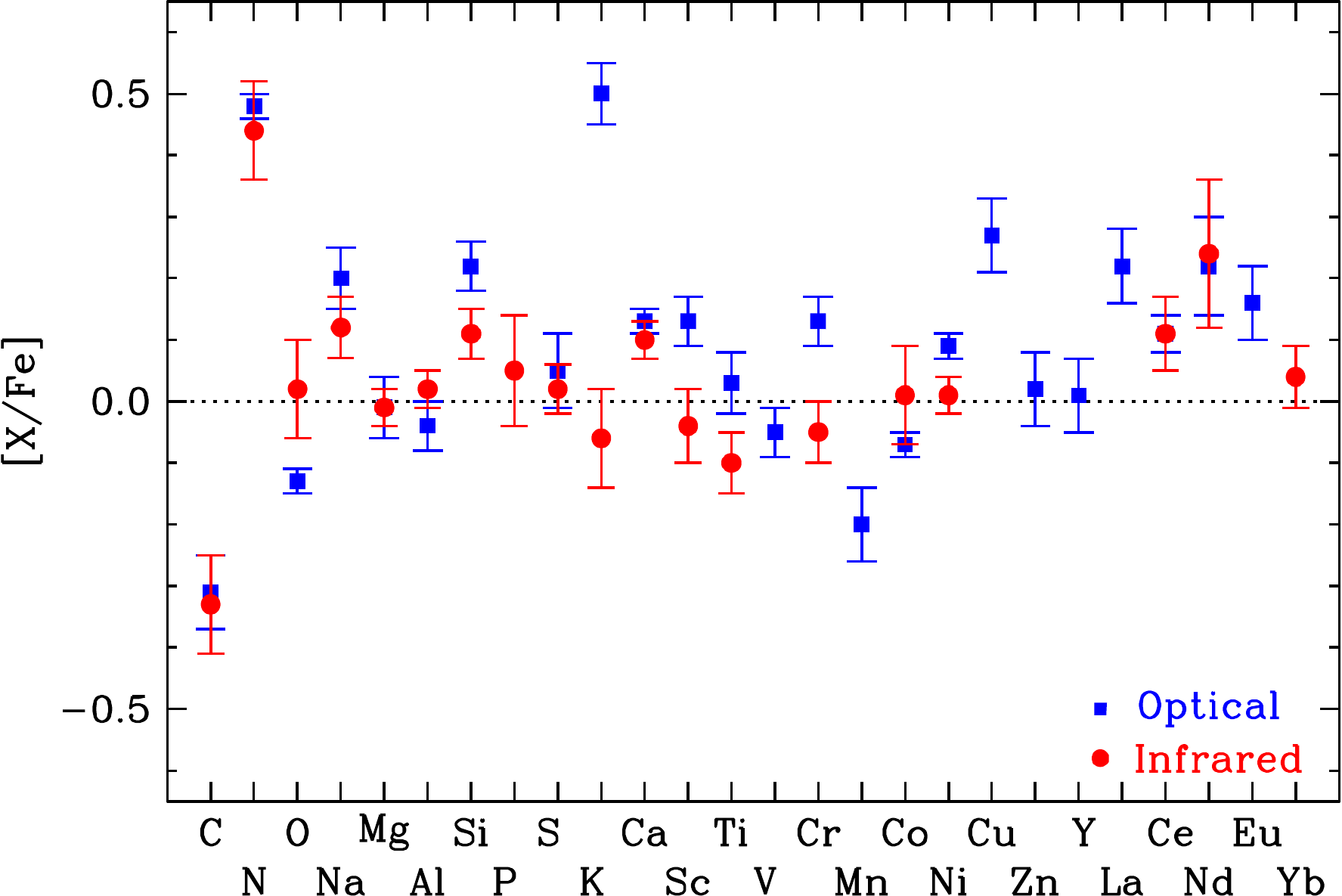}
      \caption{NGC~752 cluster mean elemental abundances from optical 
                (blue symbols) and $IR$ (red symbols) spectral region.
                The data for this figure are from Table~\ref{tab-abunds}.
                For elements represented by two species (Cr and Ti), 
                the average of the species is displayed.}
     \label{InfOpt}
\end{figure}

\begin{figure*}
  \leavevmode
      \epsfxsize=14cm
      \epsfbox{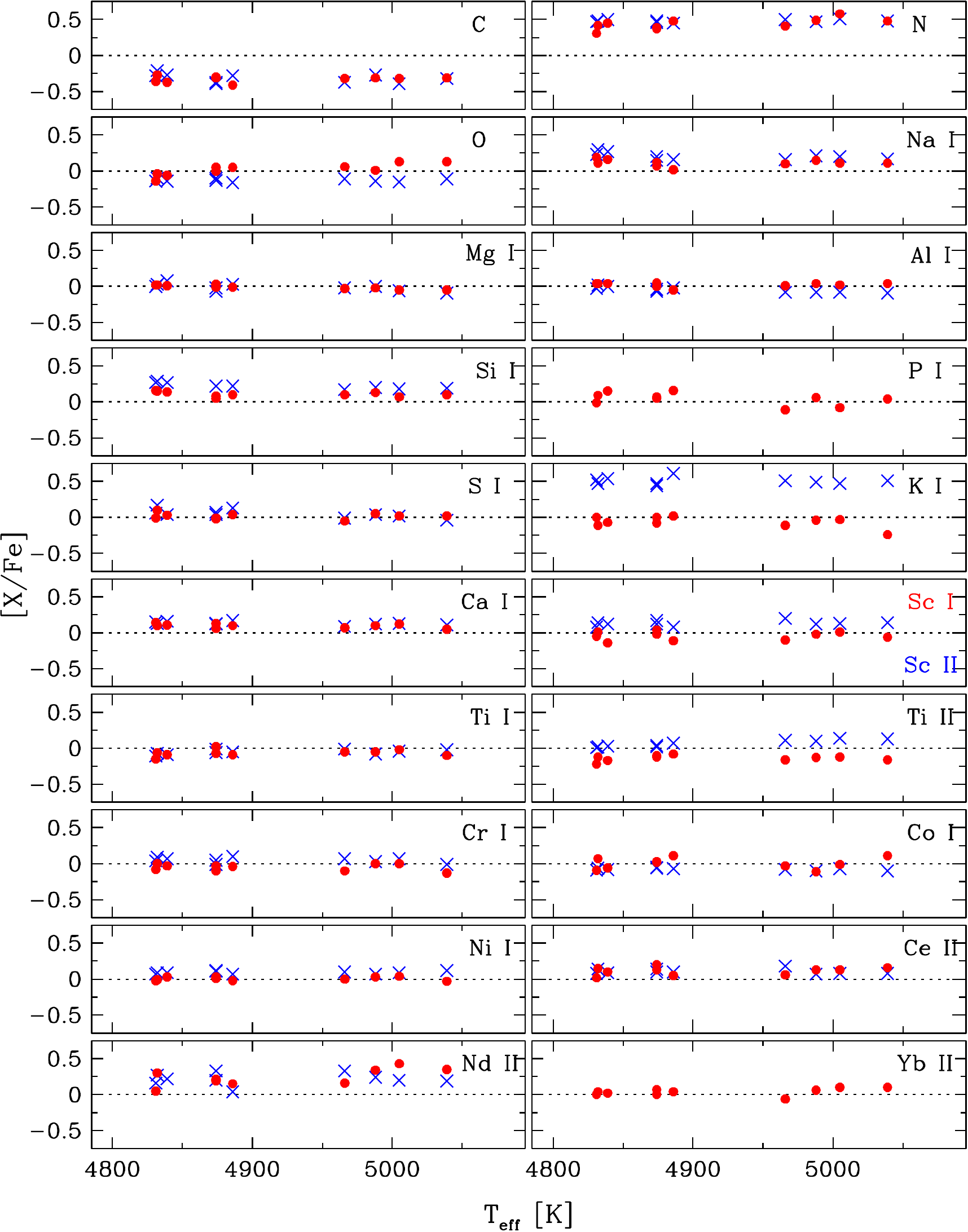}
      \caption{Mean abundances of the species for all NGC~752 program stars 
               plotted as functions of their \teff\ values.
               The panels labeled simply C, N, and O are based on multiple
               abundance indicators that are discussed in \S\ref{cnoiso}. 
               Optical and $IR$ abundances are shown with blue crosses and 
               red dots, respectively. 
               In the Sc panel, \ion{Sc}{i} (red dots) and \ion{Sc}{ii} 
               (blue crosses) represent the measurements from $IR$ and optical, 
               respectively.}
      \label{Abd}
\end{figure*}

We determined the abundances of 20 elements in NGC\,752 from the IGRINS 
$H$ and $K$ band spectra. 
Of these elements, 18 also have optical region
abundances reported by BT15.
We applied synthetic analyses to all transitions with the same atomic and 
molecular line lists and methods described in \cite{afsar18}; see Paper~1
for more detailed discussion.
We derived the abundances of H-burning (C, N, O), $\alpha$ (Mg, Si, S, Ca),
light odd-Z (Na, Al, P, K), Fe-group (Sc, Ti, Cr, Fe, Co, Ni), and 
neutron-capture (\ncap) (Ce, Nd, Yb) elements, and also determined \ciso\ 
ratios.
The relative $IR$ abundances for our NGC~752 RGs are listed in the 
second part of Table~\ref{tab-abunds}.
In Figure~\ref{InfOpt}, we plot the mean $IR$ abundances along with the  
optical ones from BT15, updated as described in \S\ref{modopt}. 
The figure shows general agreement between $IR$ and optical abundances.
Defining 
$\Delta^{\rm IR}_{\rm opt}$[A/B]~= [A/B]$_{\rm IR}$ $-$ [A/B]$_{\rm opt}$, 
we find 
$\langle\Delta^{\rm IR}_{\rm opt}$[X/Fe]$\rangle$ = 0.07~$\pm$~0.04 
($\sigma$~= 0.15)
for 18 species with both optical and $IR$ abundances.  
Figure~\ref{Abd} shows optical and $IR$ abundances of each species for 
all program RGs vs. effective temperature. 
This figure shows that in a small temperature range ($\sim$175 K) that is 
covered by our RG sample, abundances do not show significant changes with 
temperature.
Both figure~\ref{InfOpt} and \ref{Abd} indicate the 
optical/$IR$ agreement for most of the  species with a few exceptions: 
O, Sc, \ion{Ti}{ii} and \ion{K}{i}. The $IR$ abundances of these species 
deviate more than 0.15 dex from their optical counterparts. 
We will discuss these deviations in the subsections below.

\subsection{Fe-group elements:}\label{fegroup}

We have investigated NGC\,752 Fe abundances from about 20 \species{Fe}{i} 
transitions. 
As noted in our previous studies, there are no known useful \species{Fe}{ii} 
transitions in the IGRINS spectral range.
The \species{Fe}{i} transitions were adopted from \cite{afsar18}. 
In Table~\ref{tab-model} we  list the optical and $IR$ Fe abundances 
for each RG. 
Optical \species{Fe}{i} and \species{Fe}{ii} abundances were derived by 
using the $EW$ method (BT15). 
The optical cluster means are:  
$\rm \langle[\ion{Fe}{i}/H]\rangle_{\rm opt}$ $= 0.01$ ($\sigma = 0.07$) and
$\rm \langle[\ion{Fe}{ii}/H]\rangle_{\rm opt}$ $= -0.06$ ($\sigma = 0.04$). 
The 0.07 dex difference between the 
neutral and ionized iron abundances stays, in general, within the 
uncertainty limits. 
The cluster mean from the $IR$ \species{Fe}{i} lines is 
$\rm \langle[\ion{Fe}{i}/H]\rangle_{\rm IR}$ $= 0.00$ ($\sigma = 0.06$),
clearly in agreement with the optical values within the mutual uncertainties.
The metallicity of NGC~752 from the neutral- and ionized-species Fe and Ti lines,
$\rm \langle[M/H]\rangle$ $= -0.07 \pm 0.04$ (BT15), also agrees well 
with these values and indicates a solar metallicity for NGC~752.

For other Fe-group elements, we derived abundances from species \ion{Sc}{i}, 
\ion{Ti}{i}, \ion{Ti}{ii}, \ion{Cr}{i}, \ion{Co}{i} and \ion{Ni}{i}.
For Sc we used two weak $K$ band transitions, taking into account their 
hyperfine structures. 
We applied synthetic spectrum analysis to these absorption lines and 
the difference from the optical is
[$\rm \ion{Sc}{ii} / Fe]$$_{\rm opt}$ -  [$\rm \ion{Sc}{i} / Fe]$$_{\rm IR}$ = 0.12 dex. 
The difference between two spectral regions resembles the difference between 
neutral and ionized species of Cr in the 
optical and Ti both in the optical and $IR$.
We calculated Ti  abundances from 10 \ion{Ti}{i} lines and the one 
\ion{Ti}{ii} line at 15783 \AA.
Although optical and $IR$ \ion{Ti}{i} abundances are in agreement, for 
\ion{Ti}{ii} the difference is  
[$\rm \ion{Ti}{ii} / Fe]$$_{\rm opt}$ - [$\rm \ion{Ti}{ii} / 
Fe]$$_{\rm IR}$ = 0.20 dex. 
This situation was also discussed in Paper~1 and \cite{afsar18}.
For 12 RGs of NGC~6940, the difference between the optical and 
$IR$ \ion{Ti}{ii} abundances was 0.16 dex, and for the three RHB stars 
presented in \cite{afsar18} was also 0.16 dex. 
The $H$-band \ion{Ti}{ii} line comes with a CO blend but
for the temperature range for our stars its contamination of \ion{Ti}{ii} 
feature is negligible.
Further investigation of the
$IR$ \ion{Ti}{ii} line is needed to better understand the discrepancy 
between optical and $IR$ \ion{Ti}{ii} abundances.
The other Fe-group elements have agreement between optical and IR transitions.
The mean [X/Fe] abundance from Table~\ref{Abd} for Fe-group elements is
$\langle \rm [X/Fe]\rangle$ $_{\rm IR}= -0.05~(\sigma = 0.05)$ 
for six species and optical mean is 
$\langle \rm [X/Fe]\rangle$ $_{\rm opt}$ $= -0.01~(\sigma = 0.14)$ for 11 species
including \ion{V}{i}, \ion{Cr}{ii},  \ion{Mn}{i}, \ion{Cu}{i} and \ion{Zn}{i}.

\subsection{Alpha elements}\label{alphas}

\begin{figure}
\includegraphics[width=\columnwidth]{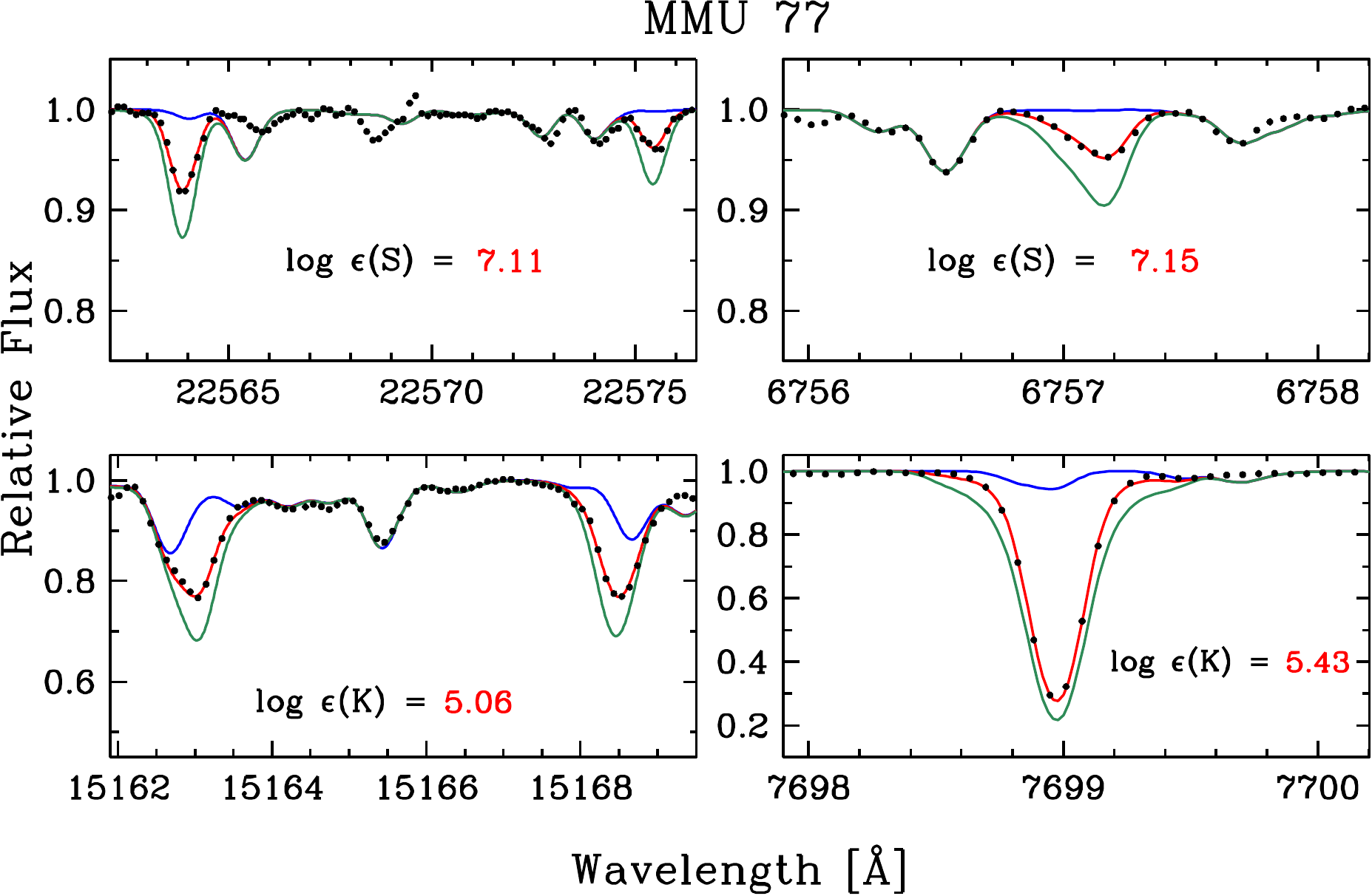}
      \caption{Observed (points) and synthetic spectra (colored lines) of 
               transitions for sulfur and potassium in $IR$ and 
               optical wavelengths.
               In each panel the blue line (in the top) represents a 
               synthesis with no contribution by the element of interest, 
               the red line (in the middle) is for the abundance that 
               best matches the observed spectrum, and the green line 
               (on the bottom) represent the synthesis 
               larger than the best fit by 0.5 dex.
  }
      \label{SK}
\end{figure}

We derived the abundances of Mg, Si, S and Ca in NGC\,752 
from their neutral species using the lines in both $H$ and $K$ bands 
(Table~\ref{Abd}).  
For the $IR$ \ion{S}{i} abundance, we made use of about ten
transitions in the $H$ and $K$ bands.
In the top panels of Figure~\ref{SK} observed and synthetic spectra of two 
\ion{S}{i} lines in the $K$ band and the combined absorption of three 
closely-spaced \ion{S}{i} lines in the optical domain for MMU~77 are shown.
Sulfur abundances are about solar for both optical and $IR$ spectral regions; 
$\langle \rm [\ion{S}{i}/Fe]\rangle_{\rm opt} = 0.05~(\sigma = 0.06)$ and 
$\langle \rm [\ion{S}{i}/Fe]\rangle_{\rm IR} = 0.02~(\sigma = 0.04)$.
Optical Ca and Si abundances from BT15 have a small line-to-line scatter 
about 0.03 dex, but Mg, on the other hand, obtained from the spectrum 
synthesis of two strong Mg lines at 5528 and 5711 \AA\ resulted in 
$\sim$0.20~dex difference. 
Mg abundances from the $IR$ region were derived from about ten absorption 
lines with a mean standard deviation of about 0.08 for 10 RGs,
which suggests greater reliability for $IR$-based Mg abundances. 
Mean abundances for $\alpha$ elements 
$\langle \rm [\alpha/Fe]\rangle \equiv \langle[Mg,Si,S,Ca/Fe]\rangle$, 
for both optical and $IR$ regions are
$\langle \rm [\alpha/Fe]\rangle_{\rm opt} = 0.10~(\sigma=0.10)$ and
$\langle \rm [\alpha/Fe]\rangle_{\rm IR} = 0.06~(\sigma=0.06)$, which
are well in agreement and slightly above solar.

\subsection{Odd-Z light elements}\label{oddz}
 
The odd-Z light elements investigated in this study 
are Na, Al and rarely-studied P and K. 
Their abundances are in Table~\ref{Abd}.
Na abundances were derived from four neutral $K$ band transitions: 
22056.4, 22083.7, 23348.4 and 23348.1 \AA. 
To our knowledge, possible non-LTE effects on these transitions have 
not yet been investigated.
The optical and $IR$ Na abundances are both above the solar values.
The $IR$ mean for NGC~752 (Table~\ref{Abd}) is 
$\langle \rm [\ion{Na}{i}/Fe]\rangle_{\rm IR} = 0.12~(\sigma=0.05)$,
while the optical mean is
$\langle \rm [\ion{Na}{i}/Fe]\rangle_{\rm opt} = 0.20~(\sigma=0.05)$. 
Al abundances were obtained from two lines in the $H$ and four lines in the $K$ band.
$H$ band abundances are always $\sim$0.1 dex lower than the $K$ band 
abundances and the lower $H$ band abundances are more in accord 
with the optical ones:
$\langle \rm [\ion{Al}{i}/Fe]\rangle_{\rm IR} = 0.02~(\sigma=0.03)$ and
$\langle \rm [\ion{Al}{i}/Fe]\rangle_{\rm opt} = -0.04~(\sigma=0.04)$.

\begin{figure}
\includegraphics[width=\columnwidth]{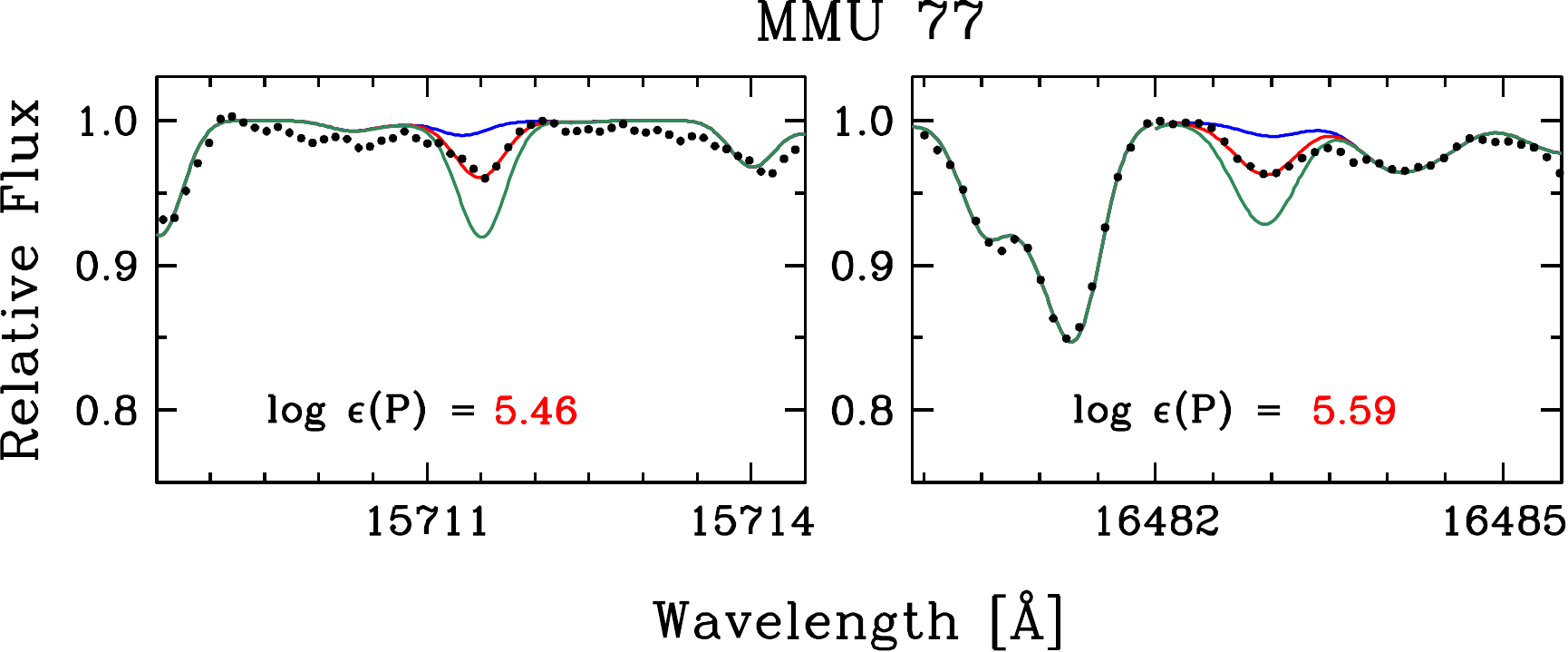}
      \caption{Observed and synthetic spectra of the phosphorus
               in the $IR$. 
               The symbols and lines have the same meanings as 
               they do in Figure~\ref{SK}.
  }
      \label{P}
\end{figure}

Phosphorus abundances were determined from two weak 
$H$-band transitions at 15711.5 and 16482.9 \AA. 
As illustrated in Figure~\ref{P} for MMU~77, the \species{P}{i} lines 
always have central depths $\lesssim$5\% for the members studied here.
The P abundance difference obtained from these two lines is 0.13 dex, but this
is an extreme case; for other program stars the agreement is much better,
usually $<$0.1~dex.
The mean phosphorus abundance,
$\langle \rm [\ion{P}{i}/Fe]\rangle_{\rm IR} = 0.04~(\sigma=0.09)$
is consistent with the solar value.

As noted in \S\ref{modopt} we derived NGC\,752 
optical-region K abundances, using the very strong \species{K}{i} resonance 
line at 7698.97 \AA\ (lower right panel of Figure~\ref{SK}). 
Our derived mean K abundance for the cluster is large,
$\langle \rm [\ion{K}{i}/Fe]\rangle_{\rm opt} = 0.50~(\sigma=0.05)$,
but this resonance line is subject to significant non-LTE effects.
\cite{takeda02} and \cite{Mucciarelli17} have computed non-LTE corrections 
between 0.2 and 0.7 dex for disk/halo stars of various \teff$-$\logg\
combinations.
\cite{afsar18} found $\sim$0.6 dex higher abundances for the 7699~\AA\ line in
three RHB stars.
Taking into account the non-LTE corrections suggested for 7698.97 \AA\
\species{K}{i} line leads to a conclusion of solar K abundance for our targets.
But since this is not based on our own calculations, we have chosen to keep 
the LTE abundance in Table~\ref{Abd}.

Potassium abundances from the $IR$ region were derived from two \ion{K}{i} lines 
at 15163.1 and 15168.4 \AA\ which are affected by CN contamination.
We illustrate this with observed/synthetic spectrum comparisons in the lower 
left panel of Figure~\ref{SK}.
Unlike the optical resonance line, the $H$ band \ion{K}{i} lines yield
approximately solar abundances:
$\rm \langle[\ion{K}{i}/Fe]\rangle_{IR} = -0.06~(\sigma=0.08)$. 
This consistency suggests that at most very small (or no) non-LTE 
corrections may be needed for these K lines.
Non-LTE studies of all detectable \ion{K}{i} lines in cool stars will be
welcome.

Following the detailed description of HF analyses in 
\cite{pilachowski15}, we have also inspected the unblended HF feature 
located at 23358.3~\AA\ in the $K$ band region. 
Unfortunately no obvious absorption of fluorine is detectable in our targets.

\subsection{\ncap\ Elements}\label{ncapels}

In this study we have obtained abundances of three \ncap\ elements 
from their ionized species transitions: Ce and Nd 
(mostly due to the \textit{s}-process in the solar-system), 
and Yb (mostly from the \textit{r}-process).
Ce abundances were derived from about four transitions, Nd from one weak 
transition at 16262.04 \AA, and Yb also from one weak line at 16498.4 \AA. 
\ion{Yb}{ii} is blended with CO but that contamination is weak enough to be 
neglected for the temperature/gravity range for our stars. 
Mean abundances of all three \ncap\ elements are about/above solar,
$\langle \rm [\ion{Ce}{ii}/Fe]\rangle_{\rm IR} = 0.11~(\sigma=0.06)$,
$\langle \rm [\ion{Nd}{ii}/Fe]\rangle_{\rm IR} = 0.24~(\sigma=0.12)$ and
$\langle \rm [\ion{Yb}{ii}/Fe]\rangle_{\rm IR} = 0.04~(\sigma=0.05)$
(Table~\ref{Abd}).
We have also analyzed optical Ce abundances from 5274.23, 5330.56, 5975.82 
and 6043.37 \AA\ transitions.
The mean value for the NGC~752 RGs is
$\langle \rm [\ion{Ce}{ii}/Fe]\rangle_{\rm opt} = 0.11~(\sigma=0.03)$, 
which is in harmony with the $IR$ abundance with smaller star-to-star scatter. 
In BT15 we derived the Nd abundances from two lines at 5255.5 and 5319.8 \AA;
the overabundance is similar what we found from the $H$ band transition, 
$\langle \rm [\ion{Nd}{ii}/Fe]\rangle_{\rm opt} = 0.22~(\sigma=0.08)$.
In Paper~1, the RGs of NGC~6940, which were analyzed in the same manner 
with the NGC~752 RGs in this study, showed slightly overabundance in 
\textit{r}-process and more in the \textit{s}-process elements. 
We observe a similar behavior; 
the simple mean of the optical and $IR$ La, Ce and Nd abundances is 
$\langle \rm [\textit{s}-process/Fe]\rangle \simeq 0.18$, 
while mean of Eu and Yb is 
$\langle \rm [\textit{r}-process/Fe]\rangle \simeq 0.10$.

\begin{figure}
\includegraphics[width=\columnwidth]{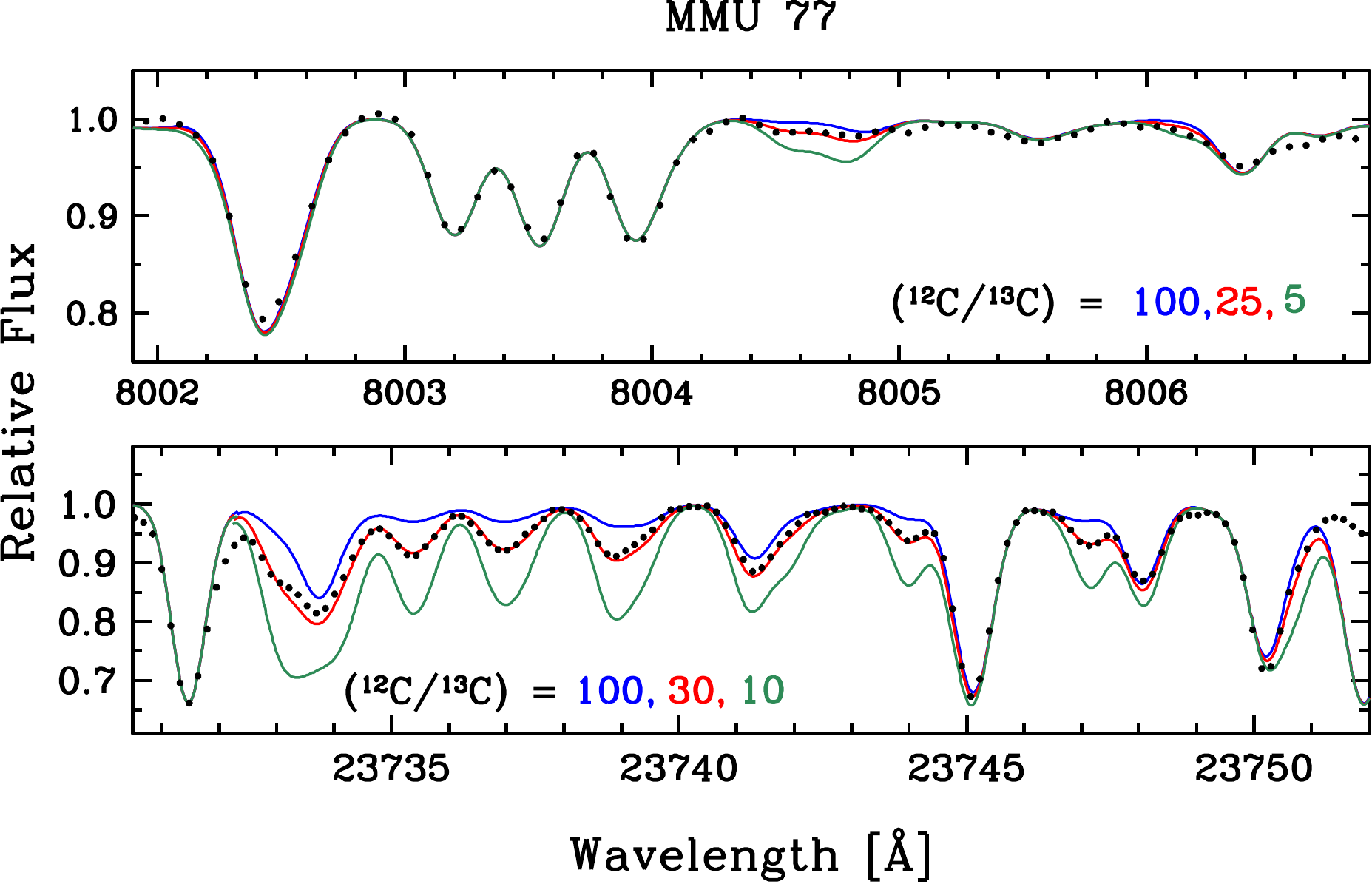}
      \caption{Observed and synthetic spectra illustrating the 
               carbon isotopic ratio of NGC~752 MMU~77 in both optical and $IR$
               regions. 
               The upper panel is centred on the triplet or \iso{13}{CN} 
               red system (2-0) lines, and the bottom panel shows the 
               \iso{13}{CO} (3-1) R-branch band head region.
               The blue, red, and green synthesis represent \ciso~=100, 
               25, and 5 in the upper panel, and \ciso~=100, 30, and 10 
               in the bottom panel, respectively.}
      \label{Ciso}
\end{figure}

\subsection{The CNO Group}\label{cnoiso}

The IGRINS spectral range contains many useful OH, CN and CO molecular 
bands that can be used to obtain CNO abundances. 
We have followed the same iterative scheme used in Paper~1
to obtain the CNO abundances.
We have also determined carbon abundances from its neutral transitions in 
both optical and $IR$ spectral regions. 

There are many OH molecular lines in the $H$ band
but most of them are very weak for our temperature and metallicity range and 
also blended with other lines and/or molecular bands. 
We were able to use about 10 OH molecular lines located between 
15200$-$17700 \AA; the abundances for each star in Table~\ref{Abd}
are simple means of the abundances derived from individual OH features. 
The resulting cluster mean is
$\langle \rm [O/Fe]\rangle_{\rm IR} = 0.02~(\sigma=0.09)$ (Table~\ref{Abd}).
In BT15, we were able use only the [\species{O}{i}] line at 6300.3 \AA\
to determine optical O abundances, and noted that this feature is plagued
with \species{Ni}{i} and CN contamination.
The calculated mean for the cluster from this line is 
$\langle \rm[O/Fe]\rangle_{\rm opt} = -0.13~(\sigma=0.02)$.
Having the advantage of obtaining O abundances from many OH features,
we rely more on the O abundance we determine from the $IR$ region.

\begin{table}
  \caption{log$\epsilon$ Abundances of Carbon in optical and infrared regions.} 
  \label{tabc}
  \begin{tabular}{@{}lccccccc@{}}
  \hline
Star & \ion{C}{i} & CH &C$_2$ & \ion{C}{i} &   CO & mean & mean \\
       &   opt        & opt           &opt      &     $IR$   & $IR$  &  opt & $IR$ \\
\hline
MMU 1	&	8.04	&	7.90	&	8.03	&	8.10	&	8.11	&	7.99	&	8.11	\\
MMU 3	&	8.16	&	7.78	&	7.98	&	8.00	&	7.98	&	7.97	&	7.99	\\
MMU 11	&	8.20	&	7.98	&	8.08	&	8.10	&	8.11	&	8.08	&	8.11	\\
MMU 24	&	8.16	&	7.80	&	7.95	&	8.00	&	8.02	&	7.97	&	8.01	\\
MMU 27	&	8.14	&	7.93	&	8.05	&	8.13	&	8.20	&	8.04	&	8.17	\\
MMU 77	&	8.05	&	7.88	&	8.05	&	8.11	&	8.21	&	7.99	&	8.16	\\
MMU 137	&	8.21	&	7.83	&	8.00	&	8.07	&	8.07	&	8.01	&	8.07	\\
MMU 295	&	8.18	&	7.95	&	8.08	&	8.11	&	8.18	&	8.07	&	8.15	\\
MMU 311	&	8.14	&	7.90	&	8.04	&	8.17	&	8.18	&	8.03	&	8.18	\\
MMU 1367&	8.18	&	7.83	&	8.03	&	8.06	&	8.03	&	8.01	&	8.05	\\
\hline
\end{tabular}
\end{table}

Carbon abundances were derived from multiple optical and $IR$ species.
The summary of the results for each star are given in Table~\ref{tabc}.
In the $IR$, we used primarily the CO molecular features in the $K$ band: 
$^{12}$CO first overtone, $\Delta$v = 2, (2-0) and (3-1) bands at 
23400 and 23700 \AA.
The scatter based on different abundance measurements for a single RG is very 
small, about $\sim$0.03 dex.
The mean C abundance from the CO molecular lines is 
$\rm \langle[C/Fe]\rangle_{CO} = -0.32~(\sigma=0.06)$, a value which would
be expected after first dredge-up and envelope mixing in metal-rich disk
stars.
There are second overtone $^{12}$CO band heads
also in the $H$ band but due to relatively high temperatures of our programme
stars they are too weak for detection.
In BT15 we obtained the carbon abundances from the CH G band, the Swan band 
heads of C$_{2}$ (0-0) at 5155 \AA\ and the (0-1) at 5635 \AA\ (Figure~9 in
BT15).
These molecular bands are heavily blended with other atomic transitions 
and the C$_{2}$ bands are weak in strength, which makes the spectral 
analysis challenging in these regions. 
But from those features BT15 derived 
$\rm \langle[C/Fe]\rangle_{CH, C_{2}} = -0.41~(\sigma=0.03)$.
Considering the analytical difficulties for CH and C$_{2}$, the $\sim$0.1 
dex difference from the $IR$ CO result indicates reasonable accord.

We obtained the carbon abundances also from the high-excitation  
\ion{C}{i} lines. 
Carbon abundances derived from the \ion{C}{i} transitions agree very well 
with CO results, 
$\rm \langle[\ion{C}{i}/Fe]\rangle_{IR} = -0.34~(\sigma=0.04)$.
As a further check we also determined the C abundances from synthetic spectrum
analyses of three high-excitation \ion{C}{i} lines located in the optical 
at 5052.1, 5380.3 and 8335.1 \AA.
The line-to-line C abundance scatter from these transitions is about 0.1 dex, 
and the mean abundance for the cluster is 
$\rm \langle[\ion{C}{i}/Fe]\rangle_{opt} = -0.22~(\sigma=0.10)$, on average 
only $\sim$0.14 dex higher compare to the mean C abundance obtained from
other features mentioned above.

In Table~\ref{tabc} we have listed the individual and mean carbon abundances.
The quoted carbon abundances in this table are the average of the 
molecular and high-excitation carbon abundances and they are in relatively 
good agreement; 
$\rm \langle[C/Fe]\rangle_{IR} = -0.33~(\sigma=0.04)$,
$\rm \langle[C/Fe]\rangle_{opt} = -0.31~(\sigma=0.06)$.

We obtained nitrogen abundances from the CN molecular 
transitions in the H-band.
We used about 18 CN features between 15000 and 15500 \AA, and calculated 
N abundances. 
The mean $IR$ N abundance is $\rm \langle[N/Fe]\rangle_{IR} = 0.44~(\sigma=0.08)$.
Optical nitrogen abundances were obtained from $^{12}$CN and $^{13}$CN 
red system lines in the 7995$-$8040 \AA\ region in BT15, and the means
are in accord with those from the $IR$,
$\rm \langle[N/Fe]\rangle_{opt} = 0.48~(\sigma=0.02)$.

Finally, we measured the \ciso\ ratios from the first 
overtone $^{12}$CO ($\Delta$v = 2) (2-0) and (3-1) band lines, which are
accompanied by the $^{13}$CO band heads near 23440 and 23730 \AA.
These are more robust features for \ciso\ ratio 
determination than the standard optical $^{13}$CN feature near 8003 \AA\ 
used by BT15.
The top panel of Figure~\ref{Ciso} shows that the $^{13}$CN triplet,
the strongest feature of this band system, is barely detectable in MMU 77
(nor is it much stronger in any NGC~752 star).
In contrast, the $^{13}$CO features shown in the bottom panel of this
figure are much stronger.
We compare the optical and $IR$ \ciso\ values in Table~\ref{tab-iso}.
They are in reasonable accord, given the extreme weakness of the CN bands.

\begin{table}
  \caption{Carbon isotopic ratios of optical and infrared regions.} 
  \label{tab-iso}
  \begin{tabular}{@{}lccc@{}}
  \hline
 Stars   &  $^{13}$CN  &  $^{13}$CO  &  $^{13}$CO \\
           & (8004 \AA) &  (23440 \AA)  &  (23730 \AA)    \\         
 \hline
MMU 1	&	25	&	25	&	30	\\
MMU 3	&	25	&	25	&	30	\\
MMU 11	&	25	&	25	&	30	\\
MMU 24	&	13	&	20	&	20	\\
MMU 27	&	17	&	19	&	25	\\
MMU 77	&	25	&	30	&	30	\\
MMU 137	&	15	&	20	&	20	\\
MMU 295	&	20	&	25	&	25	\\
MMU 311	&	15	&	23	&	22	\\
MMU 1367&	17	&	15	&	16	\\
 \hline
\end{tabular}
\end{table}

\subsection{Abundance Uncertainties}\label{uncertaities}

Detailed investigation of the internal and 
external uncertainty levels of the atmospheric parameters and their
effects on the elemental abundances were provided in BT15,
in which we calculated an average 
uncertainty limit of about 150 K by comparing the spectroscopically derived 
\teff\ values with the literature, photometric and LDR temperatures.
In Table~8 of BT15 we list the sensitivity of derived abundances 
to the model atmosphere changes within uncertainty limits for the star MMU~77.
Additional investigation of LDR temperatures from the $IR$ data
has shown that our temperature uncertainty limit has remained almost 
the same as determined in BT15, considering the highest LDR and spectral 
temperature difference of 154 K for MMU~1367 (see \S\ref{ldrcomp}).   
Therefore, in Table~\ref{temp}, we present the sensitivity of derived 
abundances in the elements only newly studied in this work adopting the same 
atmospheric parameter uncertainties in BT15. 
The uncertainties were determined using 
the $IR$ spectra of same star, MMU~77, as applied in BT15.
In general, abundance changes are mostly well within 1$\sigma$ level of 
the [X/Fe] values (Table~\ref{Abd}). 
However, the sensitivity level of \ion{Sc}{i} abundance to the change 
in temperature stands out. 
The temperature sensitivity of some $IR$ Sc lines has been previously 
noticed by \cite{thor18}, 
based on \ion{Sc}{i} lines identified in $K$ band of
cool M giants observed with NIRSPEC/Keck II.
They reported up to 0.2 dex uncertainties in Sc abundances 
mostly originated from the temperature sensitivity for stars \teff\ < 3800 K. 
Although our stars have higher temperatures and the \ion{Sc}{i} lines we used
are different than those that \citeauthor{thor18} discussed, 
caution should be taken in interpreting the $IR$ Sc
abundances for our stars until the underlying physical process for the 
temperature sensitivity of Sc lines are better understood.

\begin{table}                                                                                                                                     
 \caption{Sensitivity of elemental abundances to the model
atmosphere parameter uncertainties for MMU~77.}
 \label{temp}                                            
 \begin{tabular}{@{}lccc@{}}                             
 \hline                                                       
Species     &  $\Delta$$\teff$ (K)  &  $\Delta$\logg &  $\Delta$$\xi_{t}$ (\kmsec)  \\
                  &   $-$150 / +150 &  $-$0.25 / +0.25  &  $-$0.3 / +0.3  \\
 \hline 
\ion{P}{i}	&	0.01	/ 0.01      &	0.09	/ $-$0.09 & 0.01 / 0.00	\\
\ion{S}{i}	& $-$0.06	/ 0.09      &	0.09	/ $-$0.05 & $-$0.03 / 0.03	\\
\ion{K}{i}	&	0.10	/ $-$0.11 &	0.02	/ $-$0.01 & $-$0.01 / 0.05	\\
\ion{Sc}{i}	&	0.18	/ $-$0.18 &	0.01	/ 0.01      & 0.01 / 0.00	\\
\ion{Ce}{ii}	&	0.08	/ $-$0.06 &	0.12	/ $-$0.11 &  $-$0.03 / 0.03	\\
\ion{Yb}{ii}	&	0.05	/ $-$0.05 &	0.10	/ $-$0.10 &  $-$0.02 / 0.03	\\
 \hline
\end{tabular}                                                
\end{table}

\subsection{Comparison with\,NGC~6940}\label{comp6940}

We have now derived metallicities and relative abundance ratios for OCs
NGC\,6940 and NGC\,752 with high resolution spectra in the optical
spectral region (BT15, BT16) and infrared (BT19, this study).
The NGC\,6940 optical data were obtained with the Hobby-Eberly 
Telescope and its high-resolution echelle
spectrometer \citep{tull98}, and those for NGC\,752
with the 2.7m Smith Telescope 
and Tull echelle spectrometer \citep{tull95}; both data sets have high resolution
($R$~$\simeq$ 60,000) and high $S/N$~$\geq$~100.
The $H$ and $K$ band spectra for the two clusters were gathered with 
IGRINS set up as described in \S\ref{obs} and observed in identical 
fashions. 

\begin{figure}
\includegraphics[width=\columnwidth]{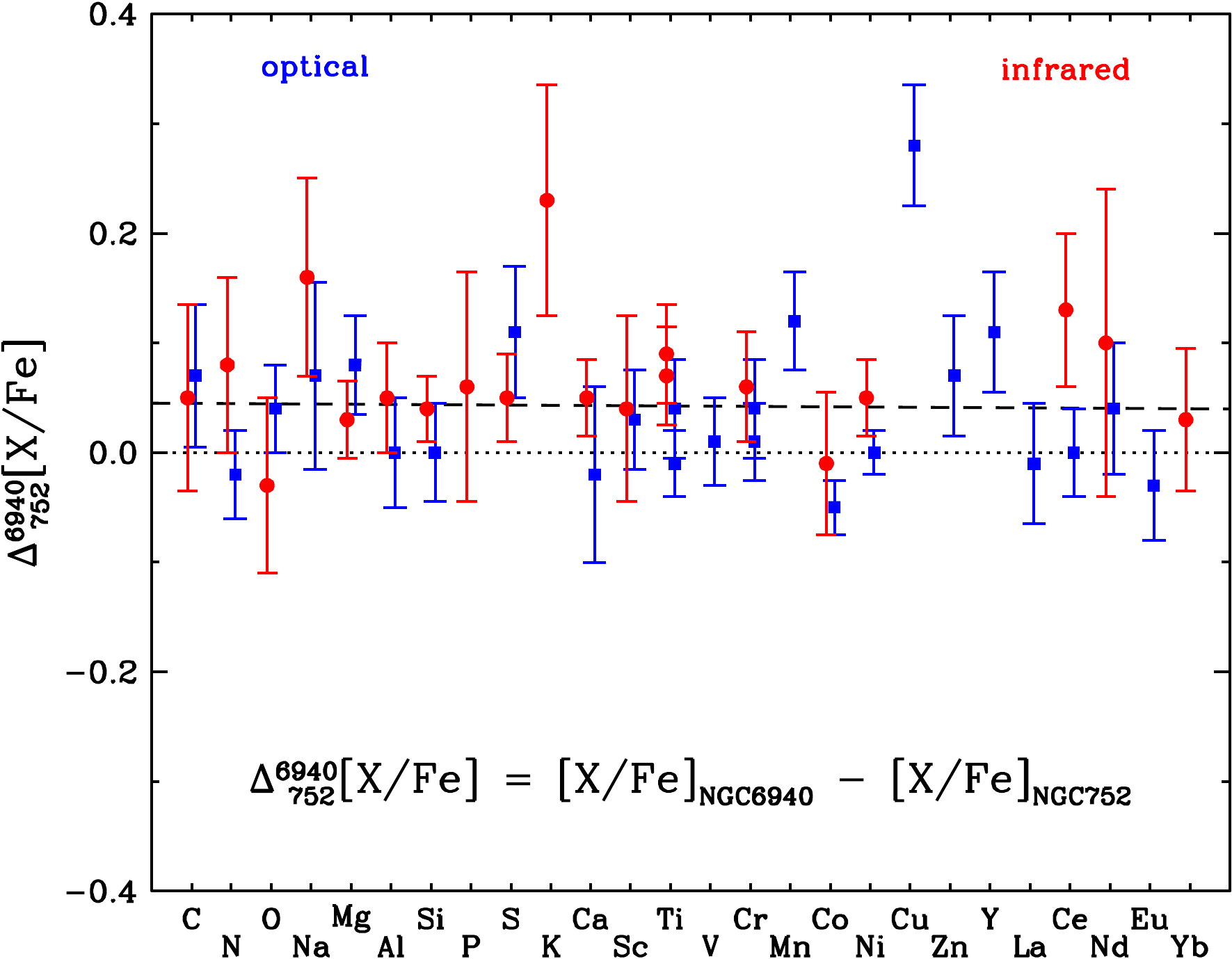}
      \caption{Differences between relative abundances [X/Fe] in NGC\,752
               and NGC\,6940 in the optical and $IR$ spectral regions.
               The dotted line at $\Delta_{ 752}^{6940}$[X/Fe] = 0.00 indicates
               equality between [X/Fe] values in the two clusters.
               The dashed line at $\Delta_{ 752}^{6940}$[X/Fe] = 0.04 
               represents the mean value for all abundances, excluding the
               aberrant values for \species{K}{i} in the $IR$ and
               \species{Cu}{i} in the optical spectral region.
      }
      \label{n752n6940}
\end{figure}

Our derived metallicities for the two clusters suggest that
NGC\,752 is slightly more metal-rich than NGC\,6940.
Defining $\Delta_{ 752}^{6940}$X = X$_{\rm NGC6940}$ $-$ X$_{\rm NGC752}$,
from optical data $\Delta_{ 752}^{6940}$[\species{Fe}{i}/H]$_{\rm opt}$ = $+$0.05 and
$\Delta_{ 752}^{6940}$[\species{Fe}{ii}/H]$_{\rm opt}$ = $+$0.08, but these differences
are well within the observational/analytical uncertainties.
The $IR$ metallicities are essentially identical:
$\Delta_{ 752}^{6940}$[\species{Fe}{i}/H]$_{\rm IR}$ = $+$0.02.
We conclude, in agreement with past studies, that both 
NGC\,6940 and NGC\,752 have solar metallicities.

The general accord between the two clusters extends to the abundance ratios
of individual elements.
In Figure~\ref{n752n6940} we show abundance differences for all species 
studied in the optical and $IR$ regions.
The uncertainties shown in the figure are approximate, being averages of 
the $\sigma$ values of the abundances in each cluster.
Excluding the aberrant points for optical \species{Cu}{i} and $IR$ 
\species{K}{i}, we derive <$\Delta_{ 752}^{6940}$[X/Fe]> = $+$0.045
(+0.06 in the optical, +0.03 in the $IR$).
The Cu difference is not well determined, as the NGC~752 
optical spectra permitted use of only one \species{Cu}{i} feature.
At present we lack an explanation for the 0.2~dex abundance difference
between the $IR$-based \species{K}{i} lines in NGC~6940 and NGC~752.
This issue will be considered again in our future studies of M67 and
other OCs. In Table~\ref{tab-diff}, we have listed the abundance differences between two clusters that generate Figure 8.
Table~\ref{tab-diff} also contains the comparison with the recent optical abundances from the literature.
The comparison with three studies \citep{carrera11,reddy12,lum19}
shows a general accord in uncertainty limits in both regions.

\begin{table}
  \caption{Abundance differences.}
   \label{tab-diff}
  \begin{tabular}{@{}lrrrr@{}}
  \hline
Species	&	N6940-N752	&	Lum - us	&	Carrera-us	&	Reddy-us	\\
  \hline
	&	\multicolumn{4}{c}{Optical Spectral Region} \\							
C	&	0.07	&	0.09	&		&		\\
N	&	$-0.02$	&	$-0.20$	&		&		\\
O	&	0.04	&	-0.02	&		&		\\
\ion{Na}{i}	&	0.07	&	-0.07	&	$-0.23$	&	$-0.08$	\\
\ion{Mg}{i}	&	0.08	&	0.06	&	$-0.03$	&	0.00	\\
\ion{Al}{i}	&	0.00	&	0.36	&	$-0.02$	&	0.19	\\
\ion{Si}{i}	&	0.00	&	$-0.11$	&	$-0.28$	&	$-0.11$	\\
\ion{S}{i}	&	0.11	&		&		&		\\
\ion{K}{i}	&		&		&		&		\\
\ion{Ca}{i}	&	$-0.02$	&	$-0.08$	&	$-0.15$	&	$-0.10$	\\
\ion{Sc}{ii}	&	0.03	&	$-0.09$	&		&	$-0.09$	\\
\ion{Ti}{i}	&	$-0.01$	&	0.11	&		&	$-0.02$	\\
\ion{Ti}{ii}	&	0.04	&	$-0.03$	&		&	$-0.11$	\\
\ion{V}{i}	&	0.01	&	0.18	&		&	0.08	\\
\ion{Cr}{i}	&	0.01	&	0.00	&	$-0.03$	&	$-0.08$	\\
\ion{Cr}{ii}	&	0.04	&	$-0.30$	&		&	$-0.22$	\\
\ion{Mn}{i}	&	0.12	&	0.24	&	0.17	&	0.07	\\
\ion{Co}{i}	&	$-0.05$	&	0.18	&		&	0.04	\\
\ion{Ni}{i}	&	0.00	&	$-0.09$	&	$-0.12$	&	$-0.10$	\\
\ion{Cu}{i}	&	0.28	&		&		&	0.16	\\
\ion{Zn}{i}	&	0.07	&	$-0.08$	&		&	$-0.12$	\\
\ion{Y}{ii}	&	0.11	&	0.04	&		&	0.02	\\
\ion{La}{ii}	&	$-0.01$	&		&		&	$-0.09$	\\
\ion{Ce}{ii}	&	0.00	&	$-0.04$	&		&	0.02	\\
\ion{Nd}{ii}	&	0.04	&	$-0.07$	&		&	$-0.16$	\\
\ion{Eu}{ii}	&	$-0.03$	&		&		&	$-0.09$	\\
average	&	0.04	&	0.00	&	$-0.09$	&	$-0.04$	\\
sigma	&	0.07	&	0.15	&	0.14	&	0.10	\\
&		&		&		\\
	&	\multicolumn{4}{c}{IGRINS H \& K  Spectral Region} \\							
C	&	0.05	&	0.11	&		&		\\
N	&	0.08	&	$-0.16$	&		&		\\
O	&	$-0.03$	&	$-0.17$	&		&		\\
\ion{Na}{i}	&	0.16	&	0.01	&	$-0.15$	&	0.00	\\
\ion{Mg}{i}	&	0.03	&	0.06	&	$-0.03$	&	0.00	\\
\ion{Al}{i}	&	0.05	&	0.30	&	$-0.08$	&	0.13	\\
\ion{Si}{i}	&	0.04	&	0.00	&	$-0.17$	&	0.00	\\
\ion{P}{i}	&	0.06	&		&		&		\\
\ion{S}{i}	&	0.05	&		&		&		\\
\ion{K}{i}	&	0.23	&		&		&		\\
\ion{Ca}{i}	&	0.05	&	$-0.05$	&	$-0.12$	&	$-0.07$	\\
\ion{Sc}{i}	&	0.04	&		&		&	0.11	\\
\ion{Ti}{i}	&	0.07	&	0.12	&		&	$-0.01$	\\
\ion{Ti}{ii}	&	0.09	&	0.18	&		&	0.10	\\
\ion{Cr}{i}	&	0.06	&	0.10	&	0.07	&	0.02	\\
\ion{Co}{i}	&	$-0.01$	&	0.10	&		&	$-0.04$	\\
\ion{Ni}{i}	&	0.05	&	$-0.01$	&	$-0.04$	&	$-0.02$	\\
\ion{Ce}{ii}	&	0.13	&	$-0.04$	&		&	0.02	\\
\ion{Nd}{ii}	&	0.10	&	$-0.09$	&		&	$-0.18$	\\
\ion{Yb}{ii}	&	0.03	&		&		&		\\
average	&	0.07	&	0.03	&	$-0.07$	&	0.00	\\
sigma	&	0.06	&	0.13	&	0.08	&	0.08	\\
 \hline
\end{tabular}
\end{table}

\section{The Age of NGC 752}\label{age}
  
In Paper I stellar evolutionary models with a solar abundance set
were fitted to the CMD of NGC$\,$6940, yielding an age of 1.15 Gyr.
To well within the uncertainties, the metallicity that we have determined for
NGC$\,$752 is the same as that of NGC$\,$6940.  
Both clusters appear to have the same helium abundance as well, given that 
(as discussed below) models for $Y=0.270$ provide equally good fits to the 
luminosities of the core He-burning red clump (RC) 
stars if (a) distance moduli based on \textit{Gaia} parallaxes are adopted,
and (b) the observed RGs in the clusters are mostly in the RC evolutionary
stage.
Inspection of the CMD for NGC~6940 (Figure~1 of Paper~1) suggests
that most RGs in that cluster are not associated with the RGB evolutionary
tracks, and thus truly are RC stars.
For NGC~752 we discuss this issue below, but for the moment simply assume
that our program stars are mostly RCs.
Then
to derive our best estimate of the age of NGC$\,$752, it is
simply a matter of interpolating in the same model grids that were used in
Paper I to identify which isochrone provides the best fit to the cluster
turnoff stars.\footnote{
See Paper I for a fairly detailed description of 
the evolutionary codes and stellar models that are used in the present 
series of papers --- including, in particular, a discussion of the treatment 
of convective core overshooting during the main-sequence (MS) phase.}

However, this process involves the cluster reddening, for which most 
estimates fall in the range $0.03 \le E(B-V) \le 0.05$ 
\citep{dan94,taylor:07,schlafly:11,twarog:15}, and the adopted 
color--\teff\ relations (from \citealt{casagrande:18}; hereafter CV18).
Fortunately, the Sun provides a valuable constraint on both the predicted
$T_{\rm eff}$~and color scales.  
According to CV18, their determinations of $M_{G, \odot} = 4.67$ and 
$(G-G_{\rm RP})_\odot = 0.49$ from reference solar spectra are accurate to 
within $\approx 0.01$\ mag.  
Encouragingly, the bolometric corrections (BCs) derived from the MARCS 
library of theoretical stellar fluxes \citep{gustafsson:08}, yield the same 
value of $M_G$ on the assumption of [Fe/H] $= 0.0$ and the solar values of 
\teff\ and \logg, but a bluer $G-G_{\rm RP}$ color by $\approx 0.01$\ mag.  
We have therefore applied a $+0.01$ mag zero-point correction to the synthetic 
colors in order that our solar model reproduces the ``observed" 
$G-G_{\rm RP}$ color of the Sun.

\begin{figure}[t]
\includegraphics[width=\columnwidth]{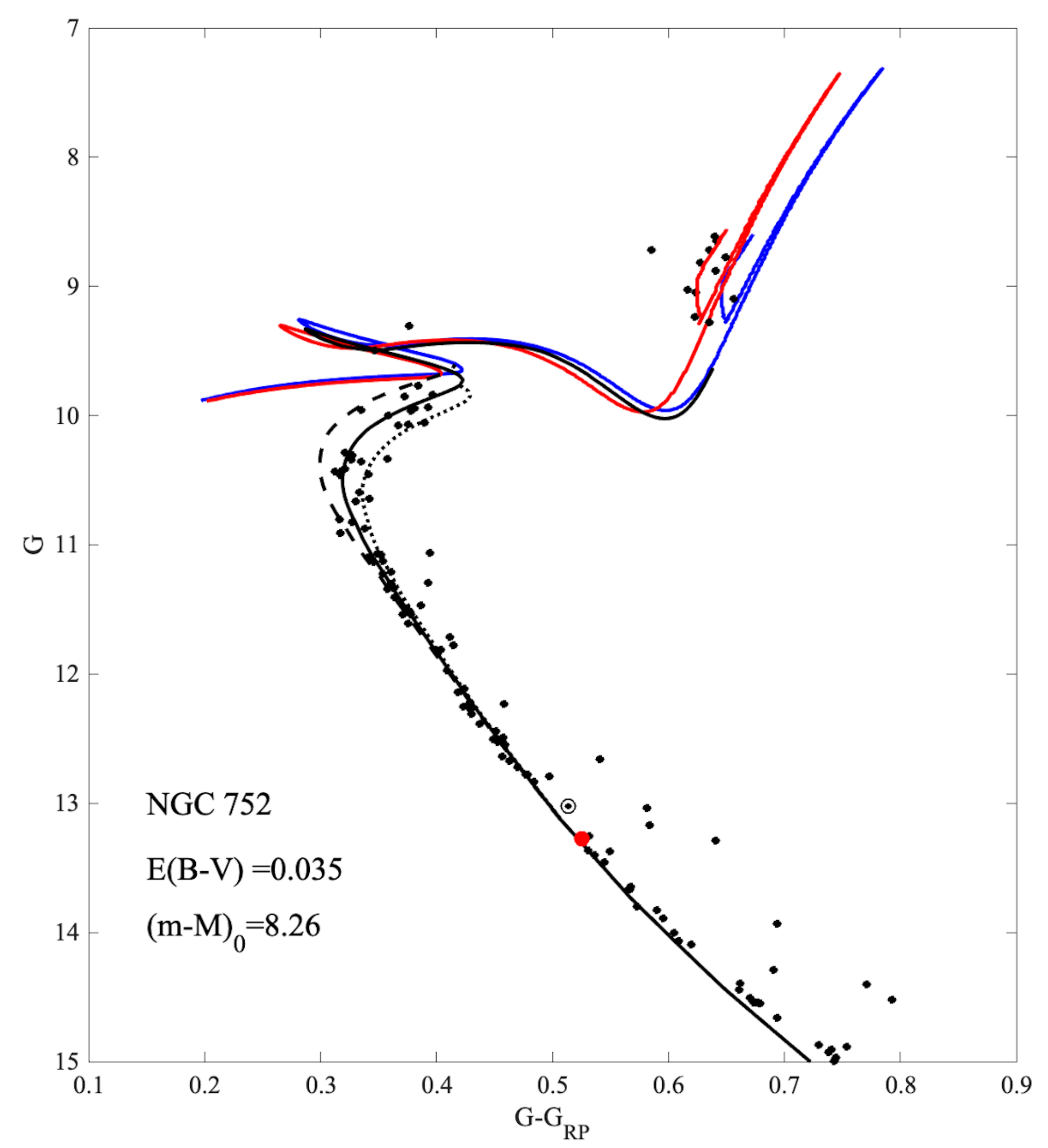}
\caption{The CMD of NGC\,752 (black dots) and its fit with a new $1.52$ Gyr 
         isochrone (solid black curve; dotted and dashed curves show 
         $1.62$ Gyr and $1.42$ Gyr isochrones). 
         The blue and red curves are the MESA evolutionary tracks computed 
         for $M=1.82\,M_\odot$, [Fe/H]\,$=0$, $Y=0.27$, assuming that 
         $f_\mathrm{ov} = 0.035$ in Equation 1 from Paper~I. 
         The red track has the mixing length increased by $10\%$ compared 
         to the solar-calibrated value of $\alpha_\mathrm{MLT} = 2$ 
         adopted for the blue track.
         See the text for descriptions of the solar symbol and the red dot.
        }
\label{fig:ngc752Gaia}
\end{figure}

In Figure~\ref{fig:ngc752Gaia} we show the best model fits 
to the observed $(G-G_{RP},\,G)$ color-magnitude diagram for NGC$\,$752.
The upper MS and turnoff stars are fit quite well by a 1.52~Gyr isochrone 
for solar abundances if the the adopted reddening is 
$E(B-V) = 0.035$\ mag.
The isochrone begins to deviate slightly to the blue of the observed MS at 
$G \sim 14$, with the offset in color at a given magnitude rising to as 
much as 0.1 mag at $G \sim 18$.
Inadequacies in the CV18 color--$T_{\rm eff}$\ relations for cooler stars are
likely responsible for this problem given that the transmission function of the
$G$ filter extends well into the ultraviolet.

The tables of BCs presented by CV18 take into account 
the dependence of the 
extinction on spectral type in a fully consistent way; i.e., these 
transformations enable one to convert predicted luminosities and 
temperatures directly to absolute $G$ and $G_{RP}$\ magnitudes that have 
been suitably corrected for an assumed reddening.
In order for the resultant models to appear on the observed $(G-G_{RP},\,G)$
CMD, they must then be shifted in the vertical direction by an amount
corresponding to the true distance modulus, $(m-M)_0$.  
In Figure~\ref{fig:ngc752Gaia} the solar symbol indicates where the Sun 
would be located if it was as distant as NGC$\,$752 and subject to the same 
reddening.  
The red filled circle represents a model at an age of 1.52~Gyr along an 
evolutionary track that has been calculated for a Standard Solar Model, and 
similarly adjusted by the adopted reddening and distance modulus.  
Thus, in order to satisfy the solar constraint, the
reddening of NGC$\,$752 must be quite close to $E(B-V) = 0.035$\ mag if it
has $(m-M)_0 = 8.26$. The inferred reddening would be larger than this if the
cluster is less distant, and vice versa.

Ages in the range of 1.7--2.0 Gyr were typically found for NGC$\,$752
in the mid-1990s \cite[e.g.][]{dan94,dinescu:95}, but
subsequent determinations have generally favored ages closer to 1.5 Gyr 
\citep[e.g.][]{anthony-twarog:06,twarog:15}.
The earlier age determinations are especially uncertain because the stellar 
models used in those studies assumed little or no overshooting from 
convective cores during the MS phase.  
Because such isochrones are incapable of reproducing the observed
turnoff morphology, the ages derived from them are highly questionable.  
In contrast, later investigations employed models that allowed for significant
amounts of core overshooting, resulting in fits to the NGC$\,$752 CMD that are
quite similar to that shown in Fig.~\ref{fig:ngc752Gaia}, and they yield
ages that have little ambiguity.

Recently, a similar color-magnitude diagram study by \cite{agueros:18}
obtained an age of $1.34 \pm 0.06$ Gyr for NGC$\,$752,
which is inconsistent with our determination by more than $2\,\sigma$.
Although those researchers used a sophisticated Bayesian approach to
determine the cluster parameters (including such observational properties
as the distance, metallicity, and extinction) from fits of isochrones
to the photometric data, their results will be subject to systematic
errors that are very difficult to quantify.  In particular, the predicted
\teff\ scale is quite sensitive to, e.g., the adopted atmospheric
boundary condition and the treatment of super-adiabatic convection.
Errors in the adopted color transformations can further impact how well
stellar models are able to reproduce an observed CMD.  Consequently, one
cannot rely on such isochrone predictions as the location of the giant
branch relative to the turnoff to provide a useful constraint on
absolute cluster ages (see, e.g., \cite{vandenberg:90}, who show that
this diagnostic may be used to obtain accurate {\it relative} ages of
star clusters.) We suspect that the derivation by \citeauthor{agueros:18} of
$A_V = 0.198\pm 0.0085$\ mag, which is appreciably higher than most
determinations, including the line-of-sight Galactic extinction
\citep{schlafly:11}, can be attributed, in part, to errors 
in the model \teff\ and/or color scales.

Isochrones appropriate to young and intermediate-age clusters
are also very dependent on how much overshooting from convective cores
during the MS phase is assumed.  In fact, the MESA models \citep{choi:16}
that were used by \citeauthor{agueros:18} assume a value of the
overshooting parameter that is, according to our analysis (see Paper I
and the next section) too low by about a factor of two.  This appears to
be the main reason (see below) why they obtained a significantly younger
age than our determination.  Unfortunately, \citeauthor{agueros:18} do not include
a figure that compares their best-fit isochrone with the CMD of
NGC$\,$752; hence it is not possible to make a visual assessment of how
well the data are fitted. Our age determination should be particularly 
robust because we have used the Sun to calibrate the predicted \teff\ 
and color scales, and have adopted the \textit{Gaia} distance and a 
spectroscopically derived metallicity, from which we have deduced 
the $E(B-V) \approx 0.035$ in order to achieve consistency with the solar constraint.  
Thus, nearly all of the cluster parameters are derived independently of our 
stellar models and the age is effectively obtained from an overlay of the 
isochrone that provides the best fit to the turnoff stars.

\section{Stellar evolution modeling of NGC\,752}\label{isoch}

In Paper~I we emphasized the importance of calibrating the efficiency of 
convective overshooting beyond the Schwarzschild boundary of the hydrogen 
convective core in MS stars with $1\la M/M_\odot\la 2$. 
In the Victoria stellar evolution code employed here to generate isochrones, 
the convective overshooting is estimated using the integral equations of 
\cite{roxburgh:89} as described in \cite{vandenberg:06}. 
In particular, Figure~1 in the latter paper shows the variation of the free 
parameter $F_\mathrm{over}$ in Roxburgh's equations calibrated by comparing 
the predicted and observed CMDs for a number of open clusters with 
different ages.
This parameter starts to increase from $F_\mathrm{over}=0$ at 
$M\approx 1.14\,M_\odot$, reaches a maximum value of $F_\mathrm{over}=0.55$ 
at $M = 1.7\,M_\odot$ and then remains constant. 
In the MESA code, that we use to model the evolution of MS turn-off (MSTO) 
stars up to the red-clump (RC) phase, the convective overshooting is 
approximated by a diffusion coefficient that is exponentially decreasing 
outside the convective boundary on a lengthscale of $0.5f_\mathrm{ov}H_P$, 
where $H_P$ is a local pressure scale height. 
In Paper~I we showed that the MESA code with $f_\mathrm{ov}=0.035$ produces 
a stellar evolution track for an initial mass $M=2\,M_\odot$ that is 
approximately equal to the MSTO mass of stars in the open cluster NGC\,6940, 
in the excellent agreement with the Victoria 1.15 Gyr isochrone generated 
for this cluster. 

\begin{figure}[t]
\includegraphics[width=\columnwidth]{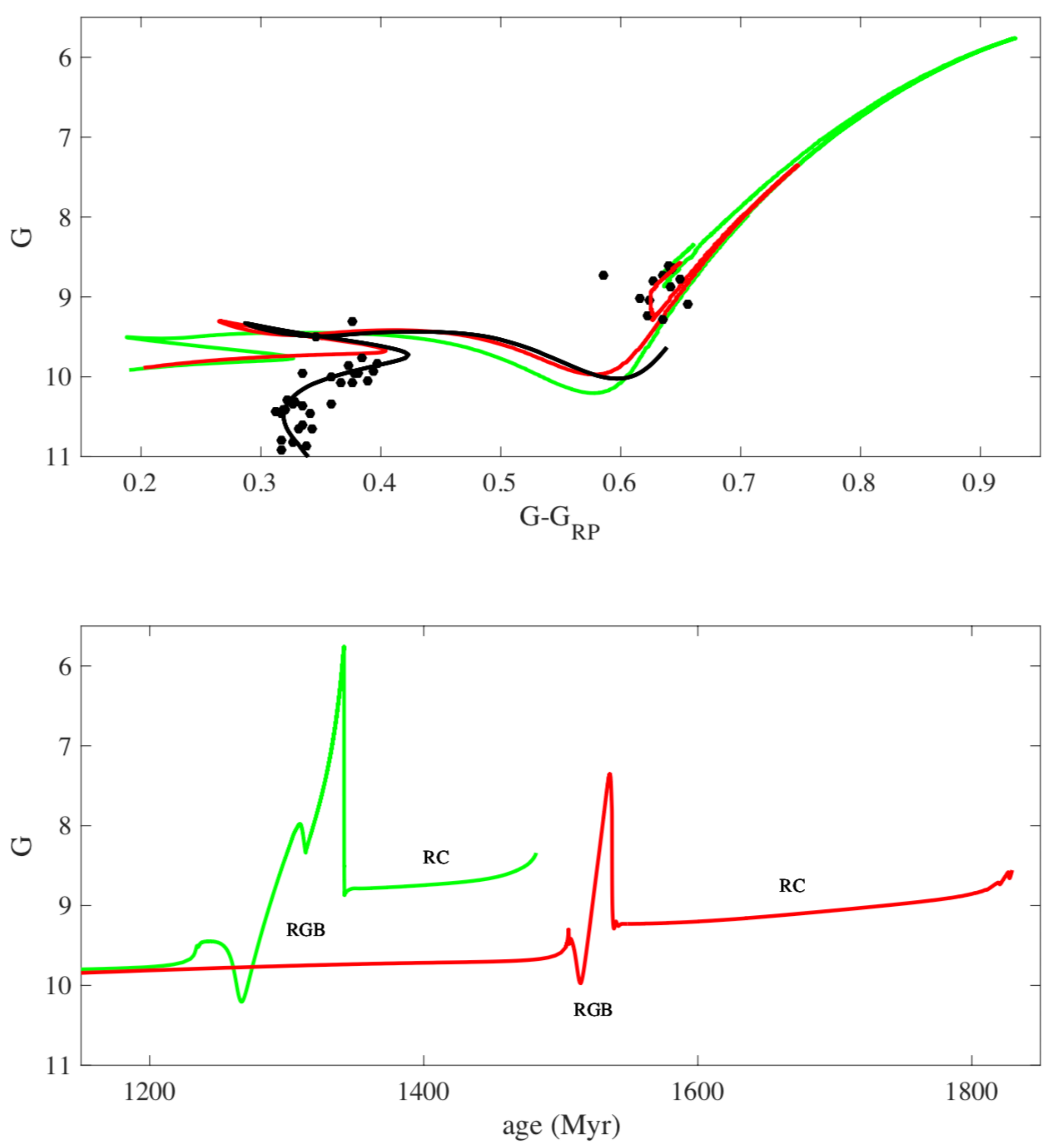}
\caption{Upper panel: the MESA evolutionary tracks of the solar-metallicity 
         $1.82\,M_\odot$ model with the convective overshooting parameter
         $f_\mathrm{ov}=0.035$ (red) and of the $1.85\,M_\odot$ model with 
         $f_\mathrm{ov}=0.016$ (green)
         that both fit the luminosity of the 
         stars leaving the MS in NGC\,752. 
         Lower panel: the RGB and RC evolutionary timescales of these models.
        }
\label{fig10new}
\end{figure}

The MSTO mass for the estimated 1.52~Gyr age of NGC\,752 is 
$M\approx 1.8\,M_\odot$.
This mass is high enough that the maximum value of $F_\mathrm{over} = 0.55$ 
should still be used according to Figure~1 in \cite{vandenberg:06}. 
Therefore we have used the same value of $f_\mathrm{ov}=0.035$ in the 
MESA code to model the evolution of MSTO stars in NGC\,752.
Figure~\ref{fig:ngc752Gaia} demonstrates that in this case the evolutionary 
track for the initial mass $1.82\,M_\odot$ 
(blue curve) is again in an 
excellent agreement with a post-MSTO part of the isochrone (black curve). 

We believe that most/all of our program stars are 
red clump (RC) members.  
First, their derived CNO abundances and \ciso\ ratios are all consistent with 
evolution beyond the first-ascent RGB, with completion of ``first dredge-up'' 
envelope mixing of CN-cyle products.
Second, stars with the NGC~752 turnoff mass should spend little time on the
upper RGB, and none are observed.
Based on the comparison of the timescales of the RGB and RC evolution for
our best-fit stellar model, we conclude that most of the red giants observed 
in NGC752 should be RC stars.

As in Paper~I, the red giant branches of the track and the isochrone are 
$\sim0.05$ mag redder than they should be for the track to be able to 
almost perfectly fit the colors of the RC stars in NGC\,752 on the following 
He-core burning evolutionary phase. 
Possible causes of this small discrepancy are mentioned in Paper~I. 
To show that our track can successfully reproduce colors and magnitudes of 
RC stars in NGC\,752, we have increased the solar-calibrated convective 
mixing length parameter $\alpha_\mathrm{MLT} = 2.0$ used for the 
red track by 10\%.
This has enhanced heat transport in the convective envelopes of our RGB 
and RC model stars and, as a result, changed their colors by 
$\sim-0.05$ (red curve in Figure~\ref{fig:ngc752Gaia}). 
We used the same remedy in Paper~I. 
It does not solve the problem of the RGB color discrepancy, but it enables 
us to see how the RGB and RC evolutionary tracks will look after the true 
cause of this discrepancy is found and fixed.

When we reduce the convective overshooting parameter 
in our solar-metallicity 
$1.82\,M_\odot$ MESA stellar models to $f_\mathrm{ov} = 0.016$ (the value
used by \citealt{choi:16} for core overshooting), and slightly increase 
the initial mass to $1.85\,M_\odot$ to keep the same MSTO luminosity, 
the morphology of their corresponding evolutionary track becomes 
inconsistent with the observed CMD of NGC\,752 
(green curves in Figure~\ref{fig10new}).
There are multiple problems: (a) the effective temperature at the end 
of the core H-burning phase is too high; (b) the track produced by core 
He-burning is too narrow in color and it does not reach the minimum 
luminosity of the observed RC stars (top panel); and (c) the RGB and RC 
evolutionary timescales are now comparable (bottom panel), which would lead
us to expect comparable numbers of RGB and RC stars in NGC\,752, 
which is not observed. 
The last inconsistency arises because the reduced efficiency of convective 
H-core overshooting leads to an extended RGB evolution with the He 
core becoming electron degenerate and experiencing a flash at the end, while
in the models with $f_\mathrm{ov}=0.035$ the He core remains non-degenerate, 
and He in the core is ignited quiescently.

\begin{figure}
\includegraphics[width=\columnwidth]{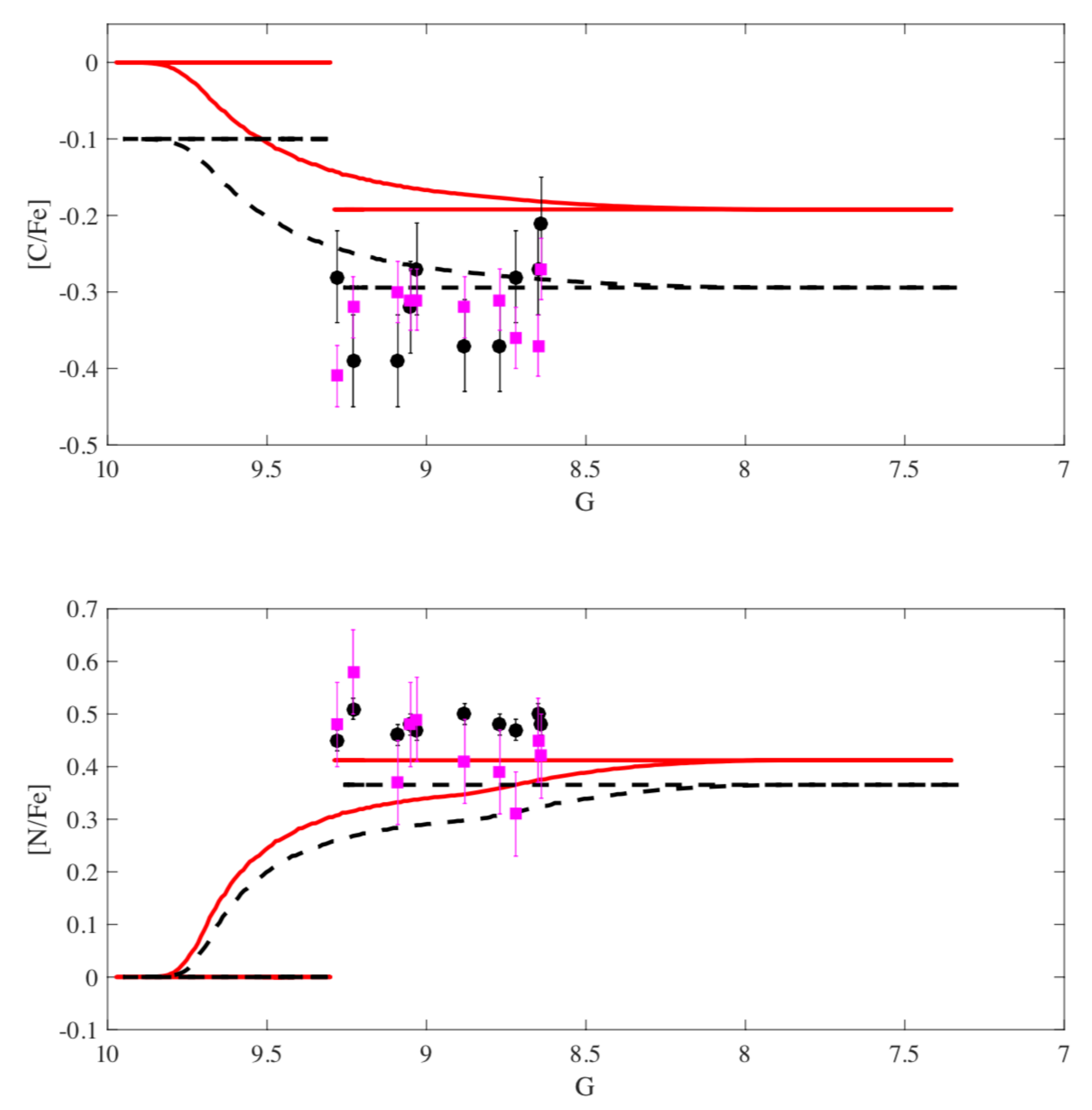}
\caption{The predicted changes of the surface C (top panel) and 
         N (bottom panel) abundances for the solar-metallicity $1.82\,M_\odot$ 
         stellar evolutionary tracks computed with $f_\mathrm{ov}=0.035$ 
         and $\alpha_\mathrm{MLT}=2.2$ are compared with the C and N 
         abundances determined for the red-clump stars in NGC\,752
         using optical (black circles) and infrared (magenta squares) spectra. 
         The red and black dashed curves
         are obtained assuming that
         the initial C abundance is [C/Fe]\,$=0$ and [C/Fe]\,$=-0.1$, respectively.
        }
\label{fig11new}
\end{figure}

Applying these evolutionary computations to C and N
abundances, in the upper panel of 
Figure~\ref{fig11new} we compare the predicted and observed 
[C/Fe] abundance ratios for the RC stars in NGC\,752. 
The observed C abundances can be reproduced by our models only if we assume 
that they were already slightly reduced initially by $\simeq$ $0.1$\,dex, 
compared to the solar-scaled [C/Fe] ratio (the dashed black curve),
because without this assumption the predicted RC abundance is [C/Fe]\,$=-0.19$.
while our mean observed value is [C/Fe]\,$=-0.31\pm 0.06$.
\cite{lum19} support a slightly subsolar initial C abundance in NGC\,752,
deriving [C/Fe]\,= $-$0.10 ($\sigma$~=0.13, 21 stars).
However, the \citeauthor{lum19} red giant abundance, [C/Fe]\,= $-$0.22 
($\sigma$~=0.08, 6 stars) is consistent with our predictions with solar or
slightly subsolar initial C abundances.

In the lower panel of Figure~\ref{fig11new} 
we make the same kind of comparison for N.
The observed optical and $IR$ values, [N/Fe]\,$=0.48\pm 0.02$ and
0.44 $\pm$ 0.08, respectively, are slightly larger than the predictions,
[N/Fe]\,= 0.41 (red curve) and 0.37 (black dashed curve)
The initial N abundance has been assumed to be solar.
However, \cite{lum19} derives [N/Fe]\,= 0.12 (no stated $\sigma$), and
that would raise the predicted red giant N abundance to be nearly comparable
to the observed one.\footnote{
The \cite{lum19} red giant abundance mean is [N/Fe]\,= 0.28 ($\sigma$ = 0.07),
somewhat lower than our predicted and observed N values.}

Finally, for carbon isotopic ratios our model predicts 
\ciso~= 22.2 and 20.9 for the red and black dashed tracks.
They are comparable with the mean C isotopic ratios measured in the RC 
stars in NGC\,752: \ciso = 22 (optical) and 16 ($IR$).
Note that our predicted C and N abundances are in a good agreement with those 
obtained for a non-rotating $1.8\,M_\odot$ star by \cite{charbonnel:10}
(\ciso\,$=19.9$, [C/Fe]\,$=-0.18$, [N/Fe]\,$=0.37$), 
who did not consider any convective overshooting, but
did include thermohaline mixing on the RGB. In our models, the enhanced 
convective overshooting significantly decreases the RGB evolution time 
(the lower panel in Figure~\ref{fig10new}), therefore if we included 
thermohaline mixing on the RGB its effect on the surface abundances of 
C and N would be even less pronounced and our assumption on the 
reduced initial abundance of C would still be required.
According to \cite{charbonnel:10}, rotation with a ZAMS velocity of 
$110\,\mathrm{km\,s}^{-1}$ only slightly changes these abundances to 
\ciso~$=15.2$,
[C/Fe]\,$=-0.19$ and [N/Fe]\,$=0.31$.

\section{Summary}\label{cocl}

This is the second of three papers that report analyses
of high-resolution optical and $IR$ spectra RG members of prominent OCs. 
In this study, we have performed the detailed chemical abundance analysis for 
10 RGs in the NGC~752 open cluster using the high-resolution 
near-$IR$ $H$ and $K$ band spectral data obtained with the IGRINS 
spectrograph. 
BT15 investigated the same RG members in the optical region from their 
high-resolution optical spectra, and here we combine data from both regions 
and explore the NGC~752 from a more complete wavelength window.

We revisited the CMD of NGC~752, investigating 
the membership assignments using \textit{Gaia} DR2 \citep{GAIA18b}. 
We applied a Gaussian mixture model to set the parallax bounds, leading to
an estimated cluster distance of 448~pc with a true distance modulus 
of $(m-M)_0 = 8.26$. 
We also remeasured the radial velocities of our targets from the $H$ and $K$ 
band spectra, finding  a cluster mean of 4.97 $\pm$ 0.24~\kmsec\
(Table~\ref{tab-motions}), which is in general agreement with both 
our previous optical RV (Paper 1) and \textit{Gaia}.

We applied LDR relations reported by \cite{fukue15} and 
estimated the $IR$-LDR effective temperatures for our targets. 
LDR temperatures obtained from both optical and $IR$ line depth ratios 
are in good agreement with the spectral temperatures within $\sim$150 K, 
indicating that this method provides reliable temperature estimations in 
the cases of lack of information from the optical region.
This encouraging result paves the way for dust-obscured open cluster 
chemical composition studies.

Adopting the model atmospheric parameters from 
Paper~1, we performed detailed abundance analysis for 20 
elements in the $H$ and $K$ band spectral regions of our targets. 
The abundances for 18 of these elements were determined 
both in the optical and $IR$ regions. 
In general, we derived the abundances of H-burning, $\alpha$, 
light odd-Z, Fe-group, and \ncap\ elements, and also determined \ciso\ 
ratios from both regions. 
In general, they are in accord with their optical counterparts and have 
abundances similar to their solar-system values.
$IR$ abundances of CNO and some $\alpha$ elements 
(such as Mg and S) were found to be more reliable compare to their optical  
counterparts due to more number of lines and regions used during the $IR$   
spectral analysis.

In some cases, small abundance differences were seen
between neutral and ionized species of the same element.
In particular there is only one \species{Ti}{ii} line present in the $H$ 
band and, compare to its optical counterparts, it seems to yield sub-solar 
abundances for our stars in general. 
Further investigation is needed to better understand this issue. 
For Sc abundances, \species{Sc}{ii} and \species{Sc}{i} lines were used in 
the optical and $IR$ regions, respectively. 
Their mean Sc cluster abundance differs by 0.12 dex. 
But only two weak \species{Sc}{i} lines in the $K$ band with hyperfine 
structures were used to determine the $IR$ Sc abundances, so we do not
regard this as a significant discrepancy.

To the best of our knowledge, P, S and K abundances have
been derived here for the first time for our targets, and all are consistent
with solar abundances in NGC~752. 
Potassium abundances obtained from two lines in the $H$ band indicate that 
these lines are likely to be less affected by non-LTE line formation 
problems than is the strong \species{K}{i} 7698.7 \AA\ resonance line. 
A similar suggestion could be made for the $IR$ Na lines.
They provide Na abundances $\sim$0.1 dex lower than the optical ones,
which might indicate that they are also less affected by non-LTE conditions. 

Five \ncap\ elements were identified in the spectra of 
NGC~752 RGs. 
The \ncap\ abundances in our stars resulted in somewhat overabundances both 
in the $s$-process and $r$-process elements, later being less slightly 
overabundant.
Encouragingly, the abundances of Ce and Nd, show agreement between their 
optical and $IR$ values.
Detection of a \species{Yb}{ii} line at 16498.4 \AA\ in the $H$ band 
provides a unique opportunity to study this element, since the strong 
resonance \species{Yb}{ii} 3694 \AA\ line occurs in a very crowded 
low-flux region of cool stars, essentially useless for abundance studies
in most solar-metallicity stars.

Analyzing CNO abundances using the many available
$IR$ CO, CN, and OH molecular features, we found cluster mean abundances 
from optical and $IR$ regions to be in reasonable agreement.
We suggest that $IR$ O abundances may provide more robust O abundances
than does the [\species{O}{i}] 6300.3 \AA\ optical line.
Our study multiple $^{12}$CO and $^{13}$CO first overtone band lines 
yields a similar endorsement: these regions provide more robust 
measurements of \ciso\ values than ones possible from the weak CN 
optical features near 8004 \AA.
Our CNO results indicate that all NGC~752 RC stars have
abundances consistent with those predicted from first dredge-up predictions,
\eg, \cite{charbonnel:10}.

We used the NGC~752 CMD to investigate the evolutionary 
states of 10 RG members, first concluding that they are at least mostly 
red clump members. 
The best evolutionary model for solar metallicity yielded core He-burning 
RC stars consistent with our stars if a helium abundance $Y$ = 0.270 is
adopted.
Isochrones fitted to the cluster turnoff yield an age of
1.52~Gyr for the reddening $E(B-V)$ = 0.035 mag and the turnoff mass 
$M=1.82\,M_\odot$ of NGC~752. 
Our light element abundance values, 
$\langle$[C/Fe]$\rangle$~$\simeq$~$-$0.32, 
$\langle$[N/Fe]$\rangle$~$\simeq$~$+$0.46, 
$\langle$[C/Fe]$\rangle$~$\simeq$~$-$0.05, 
and $\langle$\ciso$\rangle$~$\simeq$~22, are in reasonable accord with
those predicted by our MESA evolutionary models.

\section*{Acknowledgments}

This study has been supported by the US National Science Foundation (NSF, grant AST 16-16040),
and by the University of Texas Rex G. Baker, Jr. Centennial Research Endowment.
This work used the Immersion Grating Infrared Spectrometer (IGRINS) that
was developed under a collaboration between the University of Texas at
Austin and the Korea Astronomy and Space Science Institute (KASI) with
the financial support of the US National Science Foundation under grant
AST-1229522, of the University of Texas at Austin, and of the Korean GMT
Project of KASI.
These results made use of spectra obtained at the Discovery 
Channel Telescope at Lowell Observatory, and the 2.7m Smith telescope at 
McDonald Observatory.
Lowell is a private, non-profit institution dedicated to astrophysical 
research and public appreciation of astronomy and operates the DCT 
in partnership with Boston University, the University of Maryland, the 
University of Toledo, Northern Arizona University and Yale University.
We also gathered data from the European Space Agency (ESA) mission
\textit{Gaia} ($\rm https://www.cosmos.esa.int/gaia$), processed by the \textit{Gaia}
Data Processing and Analysis Consortium (DPAC,
$\rm https://www.cosmos.esa.int/web/gaia/dpac/consortium$).
Funding for the DPAC has been provided by national institutions, in
particular the institutions participating in the \textit{Gaia} Multilateral Agreement.
This research has made use of NASA's Astrophysics Data System
Bibliographic Services; the SIMBAD database and the VizieR service,
both operated at CDS, Strasbourg, France.
This research has made use of the WEBDA database, operated at the
Department of Theoretical Physics and Astrophysics of the Masaryk University,
and the VALD database, operated at Uppsala University, the
Institute of Astronomy RAS in Moscow, and the University of Vienna.

\section{APPENDIX}\label{membership_model}

The Gaussian mixture model is of the form:
\begin{equation}
    \Phi(\mu_{x_i},\mu_{y_i}, \epsilon_{x_i},\epsilon_{y_i}) = \phi_c + \phi_f
\end{equation}

\noindent where

\begin{align}
    \phi_c &= \frac{1-N_f}{2\pi\sqrt{\sigma_{c,x}^2 + \epsilon_{x_i}^2}\sqrt{\sigma_{c,y}^2 + \epsilon_{y_i}^2}\sqrt{1-\rho_c^2}}exp[-\frac{\alpha}{2(1-\rho_c^2)}]\\
    \\
    \phi_f &= \frac{N_f}{2\pi\sqrt{\sigma_{f,x}^2 + \epsilon_{x_i}^2}\sqrt{\sigma_{f,y}^2 + \epsilon_{y_i}^2}\sqrt{1-\rho_f^2}}exp[-\frac{\beta}{2(1-\rho_f^2)}]\\
\end{align}

\noindent and where
\begin{align}
    \alpha &= \frac{(\mu_{x_i} - \mu_{c,x})^2}{\sigma_{c,x}^2 + \epsilon_{x_i}^2} - \frac{2 \rho_c(\mu_{x_i} - \mu_{c,x})(\mu_{y_i} - \mu_{c,y})}{\sqrt{\sigma_{c,x}^2 + \epsilon_{x_i}^2}\sqrt{\sigma_{c,y}^2 + \epsilon_{y_i}^2}} + \frac{(\mu_{y_i} - \mu_{c,y})^2}{\sigma_{c,y}^2 + \epsilon_{y_i}^2}
    \\
    \beta &= \frac{(\mu_{x_i} - \mu_{f,x})^2}{\sigma_{f,x}^2 + \epsilon_{x_i}^2} - \frac{2 \rho_f(\mu_{x_i} - \mu_{f,x})(\mu_{y_i} - \mu_{f,y})}{\sqrt{\sigma_{f,x}^2 + \epsilon_{x_i}^2}\sqrt{\sigma_{f,y}^2 + \epsilon_{y_i}^2}} + \frac{(\mu_{y_i} - \mu_{f,y})^2}{\sigma_{f,y}^2 + \epsilon_{y_i}^2}
\end{align}

The notation for the $Gaia$ DR2 proper motion data and the 
11 model parameters is as follows:
\begin{align*}
\mu_{x_i}, \mu_{y_i} &= \text{proper motion components for $i^{th}$ star} \\
\epsilon_{x_i}, \epsilon_{y_i} &= \text{proper motion component errors for $i^{th}$ star} \\
N_f &= \text{field scaling parameter} \\
\mu_{c,x}, \mu_{c,y} &= \text{cluster center} \\
\mu_{f,x}, \mu_{f,y}& = \text{field center} \\
\sigma_{c,x}, \sigma_{c,y} &= \text{cluster std. deviations} \\
\sigma_{f,x}, \sigma_{f,y} &= \text{field std. deviations} \\
\rho_c, \rho_f &= \text{cluster and field correlation coefficients}
\end{align*}

While it is common to use ordinary maximum likelihood 
estimation to determine the parameters defining mixture models, 
as \cite{sanders71} did, we used an expectation-maximization (EM) 
machine-learning algorithm for finite mixtures as derived by \cite{dempster77}. 
We found that convergence of the model parameters using EM was more 
reliable than when applying MLE to our model. 
Central to the EM algorithm, the expectation of our complete-data 
log-likelihood function is of the form
\begin{align}
\label{eq:q_function}
     Q = \sum_{i=1}^{NST} T_{c_i}\log(\phi_{c_i}) + T_{f_i}\log(\phi_{f_i}) \textnormal{,}
\end{align}

\noindent where $NST$ is the number of total stars in 
our data set, and $\phi_{c_i}$ and $\phi_{f_i}$ are simply $\phi_c$ and 
$\phi_f$ evaluated at the $i^{th}$ star using the current parameter guesses. 
$T_{c_i}$ and $T_{f_i}$ are the conditional probabilities that 
the $i^{th}$ star belongs to the cluster or field distribution, respectively. 
They are calculated with 
$T_{c_i} = \phi_{c_i}\mathbin{/}(\phi_{c_i} + \phi_{f_i})$ and 
$T_{f_i} = \phi_{f_i}\mathbin{/}(\phi_{c_i} + \phi_{f_i})$. 
In our EM algorithm, \ref{eq:q_function} was maximized with respect to each 
of the 11 parameters numerically and the probabilities, which feed into it, 
were in turn updated. 
This process was iterated until convergence of the parameters. 
While the conditional probabilities $T_{c_i}$ and $T_{f_i}$ changed 
during the process of running the EM algorithm, the final $T_{c_i}$ 
probabilities after parameter convergence were the probabilities that we 
used for cluster membership determination.


  \bibliographystyle{mnras}{}
  \bibliography{totbib_752}

\bsp	
\label{lastpage}
\end{document}